\documentclass[apjl]{emulateapj}

\usepackage{comment}
\usepackage{amsmath}

\usepackage{lineno}

\usepackage{ifthen}

\newcommand{\forloop}[5][1]%
{%
\setcounter{#2}{#3}%
\ifthenelse{#4}%
	{%
	#5%
	\addtocounter{#2}{#1}%
	\forloop[#1]{#2}{\value{#2}}{#4}{#5}%
	}%
	{%
	}%
}%

\newcommand{\ctbd}[1]{}

\newcommand{\lc}{light curve}

\newcommand{\Lc}{Light curve}

\newcommand{\band}[1]{\ensuremath{#1}~band}

\newcommand{\kms}{\ensuremath{\rm km\,s^{-1}}}
\newcommand{\ms}{\ensuremath{\rm m\,s^{-1}}}

\newcommand{\gcmc}{\ensuremath{\rm g\,cm^{-3}}}
\newcommand{\ergscmsq}{\ensuremath{\rm erg\,s^{-1}\,cm^{-2}}}

\newcommand{\vsini}{\ensuremath{v \sin{i}}}
\newcommand{\feh}{\ensuremath{\rm [Fe/H]}}

\newcommand{\rsun}{\ensuremath{R_\sun}}
\newcommand{\msun}{\ensuremath{M_\sun}}
\newcommand{\lsun}{\ensuremath{L_\sun}}

\newcommand{\rstar}{\ensuremath{R_\star}}
\newcommand{\mstar}{\ensuremath{M_\star}}
\newcommand{\lstar}{\ensuremath{L_\star}}

\newcommand{\teffstar}{\ensuremath{T_{\rm eff\star}}}
\newcommand{\rhostar}{\ensuremath{\rho_\star}}
\newcommand{\loggstar}{\ensuremath{\log{g_{\star}}}}

\newcommand{\rpl}{\ensuremath{R_{p}}}
\newcommand{\mpl}{\ensuremath{M_{p}}}

\newcommand{\rhopl}{\ensuremath{\rho_{p}}}

\newcommand{\arstar}{\ensuremath{a/\rstar}}
\newcommand{\zrstar}{\ensuremath{\zeta/\rstar}}

\newcommand{\rjup}{\ensuremath{R_{\rm J}}}
\newcommand{\mjup}{\ensuremath{M_{\rm J}}}

\newcommand{\refsec}[1]{\mbox{\S\ \ref{sec:#1}}}

\newcommand{\reffigl}[1]{Figure~\ref{fig:#1}}
\newcommand{\refsecl}[1]{\mbox{Section \ref{sec:#1}}}

\newcommand{\reftabl}[1]{Table~\ref{tab:#1}}

\newcommand{\hatcurhtr}{HATS610-018}                                   
\newcommand{\hatcurCCra}{\ensuremath{11^{\mathrm h}35^{\mathrm m}49.92{\mathrm s}}}                                  
\newcommand{\hatcurCCdec}{\ensuremath{-29{\arcdeg}09{\arcmin}21.6{\arcsec}}}                                 
\newcommand{\hatcurCCtwomass}{2MASS~11354977-2909216}                  
\newcommand{\hatcurCCgsc}{GSC~6664-00410}                              
\newcommand{\hatcurCCtassmv}{\ensuremath{14.067\pm0.040}}              
\newcommand{\hatcurCCtassmB}{\ensuremath{14.870\pm0.060}}              
\newcommand{\hatcurCCtassmg}{\ensuremath{14.407\pm0.020}}              
\newcommand{\hatcurCCtassmr}{\ensuremath{13.854\pm0.030}}              
\newcommand{\hatcurCCtassmi}{\ensuremath{13.77\pm0.15}}                
\newcommand{\hatcurCCtwomassJmag}{\ensuremath{12.736\pm0.026}}         
\newcommand{\hatcurCCtwomassHmag}{\ensuremath{12.382\pm0.028}}         
\newcommand{\hatcurCCtwomassKmag}{\ensuremath{12.289\pm0.028}}         
\newcommand{\hatcurLCrprstar}{\ensuremath{0.1347\pm0.0019}}            
\newcommand{\hatcurLCbsq}{\ensuremath{0.085_{-0.054}^{+0.110}}}        
\newcommand{\hatcurLCimp}{\ensuremath{0.29_{-0.11}^{+0.15}}}           
\newcommand{\hatcurLCzeta}{\ensuremath{29.09_{-0.19}^{+0.26}}}         
\newcommand{\hatcurLCdur}{\ensuremath{0.07886\pm0.00083}}              
\newcommand{\hatcurLCingdur}{\ensuremath{0.0101\pm0.0010}}             
\newcommand{\hatcurLCP}{\ensuremath{0.83784340\pm0.00000047}}          
\newcommand{\hatcurLCPprec}{\ensuremath{0.8378434}}                    
\newcommand{\hatcurLCPshort}{\ensuremath{0.8378}}                      
\newcommand{\hatcurLCT}{\ensuremath{2457089.90598\pm0.00026}}          
\newcommand{\hatcurLCrho}{\ensuremath{1.38_{-0.21}^{+0.13}}}           
\newcommand{\hatcurSMEiteff}{\ensuremath{5700\pm140}}                  
\newcommand{\hatcurSMEizfeh}{\ensuremath{0.280\pm0.090}}               
\newcommand{\hatcurSMEizfehshort}{\ensuremath{0.28}}                   
\newcommand{\hatcurSMEilogg}{\ensuremath{4.65\pm0.15}}                 
\newcommand{\hatcurSMEivsin}{\ensuremath{5.72\pm0.65}}                 
\newcommand{\hatcurSMEivmac}{\ensuremath{0.0}}                         
\newcommand{\hatcurSMEivmic}{\ensuremath{0.0}}                         
\newcommand{\hatcurSMEiiteff}{\ensuremath{5600\pm120}}                 
\newcommand{\hatcurSMEiizfeh}{\ensuremath{0.280\pm0.080}}              
\newcommand{\hatcurSMEiizfehshort}{\ensuremath{0.28}}                  
\newcommand{\hatcurSMEiilogg}{\ensuremath{4.397\pm0.056}}              
\newcommand{\hatcurSMEiivsin}{\ensuremath{6.23\pm0.47}}                
\newcommand{\hatcurLBii}{\ensuremath{0.3097}}                          
\newcommand{\hatcurLBiii}{\ensuremath{0.3143}}                         
\newcommand{\hatcurLBir}{\ensuremath{0.4124}}                          
\newcommand{\hatcurLBiir}{\ensuremath{0.2959}}                         
\newcommand{\hatcurISOm}{\ensuremath{1.037\pm0.047}}                   
\newcommand{\hatcurISOmlong}{\ensuremath{1.037\pm0.047}}               
\newcommand{\hatcurISOr}{\ensuremath{1.020_{-0.031}^{+0.057}}}         
\newcommand{\hatcurISOrlong}{\ensuremath{1.020_{-0.031}^{+0.057}}}     
\newcommand{\hatcurISOrho}{\ensuremath{1.37_{-0.23}^{+0.12}}}          
\newcommand{\hatcurISOlogg}{\ensuremath{4.436\pm0.034}}                
\newcommand{\hatcurISOlum}{\ensuremath{0.93\pm0.13}}                   
\newcommand{\hatcurISOmv}{\ensuremath{4.94\pm0.17}}                    
\newcommand{\hatcurISOage}{\ensuremath{4.2\pm2.2}}                     
\newcommand{\hatcurISOMK}{\ensuremath{3.281\pm0.099}}                  
\newcommand{\hatcurISOspec}{G}                                         
\newcommand{\hatcurRVK}{\ensuremath{415.2\pm10.0}}                     
\newcommand{\hatcurRVgamma}{\ensuremath{7663.3\pm7.7}}                 
\newcommand{\hatcurRVjittertwosiglim}{\ensuremath{<11.4}}              
\newcommand{\hatcurPPi}{\ensuremath{85.5_{-2.8}^{+1.9}}}               
\newcommand{\hatcurPPlogg}{\ensuremath{3.435_{-0.063}^{+0.035}}}       
\newcommand{\hatcurPPar}{\ensuremath{3.71_{-0.22}^{+0.11}}}            
\newcommand{\hatcurPParel}{\ensuremath{0.01761\pm0.00027}}             
\newcommand{\hatcurPPrho}{\ensuremath{1.02_{-0.20}^{+0.13}}}           
\newcommand{\hatcurPPm}{\ensuremath{1.980\pm0.077}}                    
\newcommand{\hatcurPPmlong}{\ensuremath{1.980\pm0.077}}                
\newcommand{\hatcurPPr}{\ensuremath{1.337_{-0.049}^{+0.102}}}          
\newcommand{\hatcurPPrlong}{\ensuremath{1.337_{-0.049}^{+0.102}}}      
\newcommand{\hatcurPPmrcorr}{\ensuremath{0.36}}                        
\newcommand{\hatcurPPteff}{\ensuremath{2060\pm59}}                     
\newcommand{\hatcurPPtheta}{\ensuremath{0.0498_{-0.0033}^{+0.0025}}}   
\newcommand{\hatcurPPfluxavg}{\ensuremath{4.07\pm0.48}}                
\newcommand{\hatcurPPfluxavgdim}{\ensuremath{9}}                       
\newcommand{\hatcurXAv}{\ensuremath{0.076_{-0.076}^{+0.114}}}          
\newcommand{\hatcurXdistred}{\ensuremath{645_{-25}^{+36}}}             
\newcommand{\hatcurCCpmra}{\ensuremath{2.7\pm1.2}}                     
\newcommand{\hatcurCCpmdec}{\ensuremath{-4.4\pm1.2}}                   
\newcommand{\hatcurRVecceneccen}{\ensuremath{0.063\pm0.049}}           
\newcommand{\hatcurRVeccentwosiglimeccen}{\ensuremath{<0.166}}         
\newcommand{\hatcur}{HATS-18}
\newcommand{\hatcurb}{HATS-18b}
\newcommand{\hatcurRVgammaabs}{\hatcurRVgamma}                           

\newcommand{\hatcurlumind}{\rhostar}
\newcommand{\hatcurjhkfilset}{ESO}

\newcommand{\hatcurSMEversion}{ii}                                       
\newcommand{\hatcurSMEteff}{\ifthenelse{\equal{\hatcurSMEversion}{i}}{\hatcurSMEiteff}{\hatcurSMEiiteff}}
\newcommand{\hatcurSMEzfeh}{\ifthenelse{\equal{\hatcurSMEversion}{i}}{\hatcurSMEizfeh}{\hatcurSMEiizfeh}}
\newcommand{\hatcurSMEzfehshort}{\ifthenelse{\equal{\hatcurSMEversion}{i}}{\hatcurSMEizfehshort}{\hatcurSMEiizfehshort}}
\newcommand{\hatcurSMElogg}{\ifthenelse{\equal{\hatcurSMEversion}{i}}{\hatcurSMEilogg}{\hatcurSMEiilogg}}
\newcommand{\hatcurSMEvsin}{\ifthenelse{\equal{\hatcurSMEversion}{i}}{\hatcurSMEivsin}{\hatcurSMEiivsin}}
\newcommand{\hatcurSMEvmac}{\ifthenelse{\equal{\hatcurSMEversion}{i}}{\hatcurSMEivmac}{\hatcurSMEiivmac}}
\newcommand{\hatcurSMEvmic}{\ifthenelse{\equal{\hatcurSMEversion}{i}}{\hatcurSMEivmic}{\hatcurSMEiivmic}}

\newboolean{emulateapj}
\setboolean{emulateapj}{true}

\newboolean{rvtablelong}
\setboolean{rvtablelong}{true}

\newboolean{astroph}
\setboolean{astroph}{true}

\graphicspath{{img/}}

\makeatletter
\providecommand*{\input@path}{}
\edef\input@path{{data/}\input@path}
\makeatother

\shortauthors{Penev et al.}

\shorttitle{\hatcur\lowercase{b}}

\ifthenelse{\boolean{emulateapj}}{
    \newcommand{\titledag}{$\dagger$}
}{
    \newcommand{\titledag}{\dagger}
}

\begin{document}
\title{
\hatcur\lowercase{b}: An Extreme Short--Period Massive Transiting Planet
Spinning Up Its Star \altaffilmark{\titledag}
}

\author{
    K.~Penev\altaffilmark{1},
	J.~D.~Hartman\altaffilmark{1},
    G.~\'A.~Bakos\altaffilmark{1,$\star$,$\star\star$},
	S.~Ciceri\altaffilmark{6},
	R.~Brahm\altaffilmark{4,5}, 
	D.~Bayliss\altaffilmark{2,3}, 
    J.~Bento\altaffilmark{2},
	A.~Jord\'an\altaffilmark{4,5},
	Z.~Csubry\altaffilmark{1},  
	W.~Bhatti\altaffilmark{1},
	M.~de~Val-Borro\altaffilmark{1}, 
	N.~Espinoza\altaffilmark{4,5},
	G.~Zhou\altaffilmark{1}, 
	L.~Mancini\altaffilmark{6},
	M.~Rabus\altaffilmark{4,6},
	V.~Suc\altaffilmark{4},
	T.~Henning\altaffilmark{6},
	B.~Schmidt\altaffilmark{2},
	R.~W.~Noyes\altaffilmark{9},
	J.~L\'az\'ar\altaffilmark{12},
	I.~Papp\altaffilmark{12},
	P.~S\'ari\altaffilmark{12},
}
\altaffiltext{1}{
    Department of Astrophysical Sciences, Princeton University, NJ 08544, USA
} 
\altaffiltext{$\star$}{Alfred P.~Sloan Research Fellow}
\altaffiltext{$\star\star$}{Packard Fellow}
\altaffiltext{2}{
    Research School of Astronomy and Astrophysics,
    Australian National University, Canberra, ACT 2611, Australia
}
\altaffiltext{3}{
    Observatoire
    Astronomique de l'Universit\'e de Gen\`eve, 51 ch. des Maillettes,
    1290 Versoix, Switzerland
}
\altaffiltext{4}{
    Instituto de Astrof\'isica, Facultad de F\'isica,
    Pontificia Universidad Cat\'olica de Chile, Av. Vicu\~na Mackenna
    4860, 7820436 Macul, Santiago, Chile; rbrahm@astro.puc.cl
}
\altaffiltext{5}{Millennium Institute of Astrophysics, Av. Vicu\~na
  Mackenna 4860, 7820436 Macul, Santiago, Chile}
\altaffiltext{6}{Max
  Planck Institute for Astronomy, Heidelberg, Germany}
\altaffiltext{9}{Harvard-Smithsonian Center for Astrophysics,
  Cambridge, MA 02138, USA}
\altaffiltext{12}{Hungarian Astronomical Association, Budapest, Hungary}

\altaffiltext{$\dagger$}{
The HATSouth network is operated by a collaboration consisting of Princeton
University (PU), the Max Planck Institute f\"ur Astronomie (MPIA), the
Australian National University (ANU), and the Pontificia Universidad
Cat\'olica de Chile (PUC).  The station at Las Campanas Observatory (LCO) of
the Carnegie Institute is operated by PU in conjunction with PUC, the station
at the High Energy Spectroscopic Survey (H.E.S.S.) site is operated in
conjunction with MPIA, and the station at Siding Spring Observatory (SSO) is
operated jointly with ANU.
This paper includes data gathered with
%
%
the MPG~2.2\,m
telescope at the ESO Observatory in La Silla.  This
paper uses observations obtained with facilities of the Las Cumbres
Observatory Global Telescope.
}


\begin{abstract}

\setcounter{footnote}{10}
We report the discovery by the HATSouth network of \hatcurb{}: a
\hatcurPPm\,\mjup, \hatcurPPr\,\rjup{} planet in a \hatcurLCPshort\,day
orbit, around a solar analog star (mass \hatcurISOm\,\msun, and radius
\hatcurISOr\,\rsun) with $V=\hatcurCCtassmv$\,mag. The high planet mass,
combined with its short orbital period, implies strong tidal coupling between
the planetary orbit and the star. In fact, given its inferred age, \hatcur{}
shows evidence of significant tidal spin up, which together with WASP-19 (a
very similar system) allows us to constrain the tidal quality factor for
Sun--like stars to be in the range $6.5 \lesssim \log_{10}(Q^*/k_2) \lesssim
7$ even after allowing for extremely pessimistic model uncertainties. In
addition, the \hatcur{} system is among the best systems (and often the best
system) for testing a multitude of star--planet interactions, be they
gravitational, magnetic or radiative, as well as planet formation and
migration theories.

\setcounter{footnote}{0}
\end{abstract}

\keywords{
    planetary systems ---
    stars: individual (\hatcur) ---
    techniques: spectroscopic, photometric
}


\section{Introduction}
\label{sec:introduction}

Hot Jupiters, gas giant planets with orbital periods shorter than a few days,
are among the easiest extrasolar planets to detect through either transit or
radial velocity (RV) searches (to date, the two most productive methods).  In
spite of that, the sample of these planets is rather small, showing that they
are intrinsically rare. Among those, giant planets with extreme short-period
orbits, say under one day, are the easiest to detect yet the most scarce. In
fact, out of the 4696 candidate planet Kepler objects of interest (KOI) on
the NASA exoplanet
archive\footnote{\url{http://exoplanetarchive.ipac.caltech.edu/}}, only 229
have a radius of at least 6 Earth radii (approximately half the radius of
Jupiter) and orbital periods shorter than 5 days, and of those, only 41 have
a periods shorter than 1 day. This, combined with the fact, that these are
expected to be the KOIs with the highest chance of being false positives
\citep[c.f.][]{Fressin_et_al_13}, have the highest probability to transit and
that none of the transiting ones should be missed by Kepler, demonstrates how
unusual these planetary systems are.

On the other hand, this exotic population of planets, especially the ones
transiting their star, is very valuable, since it pushes theories of planet
formation, structure and evolution, as well as planet--star interactions to
the limit \citep[c.f.][]{Ida_Lin_08, Dawson_MurrayClay_13, Albrecht_et_al_12,
Ginzburg_Sari_15, Penev_et_al_12}. In addition, the deep and frequent
transits and large RV signals of these objects make them the easiest to carry
follow--up studies on, thus enhancing their power to constrain theories even
further.

We report the discovery by the HATSouth transit survey
\citep{bakos:2013:hatsouth} of \hatcurb{}: a very short period
(\hatcurLCPshort\,day) massive (\hatcurPPm\,\mjup) extrasolar planet around a
star very similar to our Sun (mass \hatcurISOm\,\msun, radius
\hatcurISOr\,\rsun{} and effective temperature \hatcurSMEteff\,K). Due to the
proximity of the planet to its host star, this system provides one of the best
laboratories for testing theories of star--planet interactions and planet
formation. In fact, we argue that \hatcur{} shows signs of being tidally
spun--up by the planet, and that modelling this effect for this system alone
constrains the tidal dissipation efficiency of the host star to better than an
order of magnitude even with very generous assumptions on possible formation
scenarios or model parameter uncertainties. Further, we show that expanding
such models to the few other very short period systems, should drastically
improve that constraint. Further, such modelling may begin to disentangle some
of the very poorly understood physics behind tidal dissipation by measuring its
dependence on various system properties.

The layout of the paper is as follows: in \refsec{obs} we describe the
discovery and follow--up observations used to confirm \hatcurb{} as a planet;
in \refsec{analysis} we outline the combined photometric and spectroscopic
analysis performed and give the inferred system properties; in
\refsec{comparison} we place \hatcur{} in the context of other extremely
short period exoplanet systems; in \refsec{Q} we derive constraints on the
tidal quality factor for stars similar to the Sun by modelling \hatcur{} and
WASP-19's orbital and stellar spin evolution; and we conclude with a
discussion in \refsec{discussion}.

\section{Observations}
\label{sec:obs}

\subsection{Photometry}
\label{sec:photometry}

\subsubsection{Photometric detection}
\label{sec:detection}

\begin{figure}[] \plotone{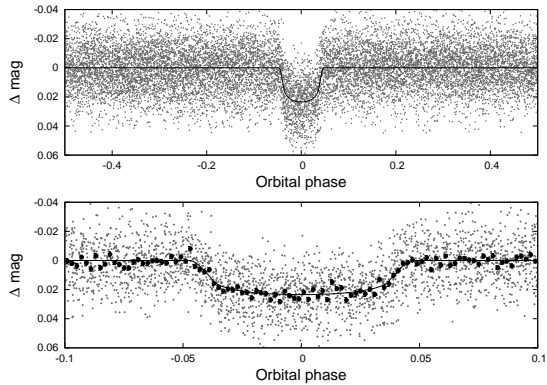} \caption[]{
        Unbinned instrumental \band{r} \lc{} of \hatcur{} folded with the
        period $P = \hatcurLCPprec$\,days resulting from the global fit
        described in \refsecl{analysis}.  The solid line shows the best-fit
        transit model (see \refsecl{analysis}).  In the lower panel we
        zoom--in on the transit; the dark filled points here show the light
        curve binned in phase using a bin-size of 0.002.\\
\label{fig:hatsouth}} \end{figure}

The star \hatcur{} (Table~\ref{tab:stellar}) was observed by HATSouth
instruments between UT 2011 April 18 and UT 2013 July 21 using the HS-2,
HS-4, and HS-6 units at the Las Campanas Observatory in Chile, the High
Energy Spectroscopic Survey site in Namibia, and Siding Spring Observatory in
Australia, respectively. A total of 5372, 3758 and 4008 images of \hatcur{}
were obtained with HS-2, HS-4 and HS-6, respectively. The observations were
obtained through a Sloan $r$ filter with an exposure time of 240\,s. The data
were reduced to trend-filtered light curves using the aperture photometry
pipeline described by \citet{penev:2013:hats1} and making use of External
Parameter Decorrelation \citep[EPD;][]{bakos:2010:hat11} and the Trend
Filtering Algorithm \citep[TFA;][]{kovacs:2005:TFA} to remove systematic
variations.  We searched for transits using the Box Least Squares
\citep[BLS;][]{kovacs:2002:BLS} algorithm, and detected a
$P=\hatcurLCPshort{}$\,day periodic transit signal in the light curve of
\hatcur{} (Figure~\ref{fig:hatsouth}; the data are available in
Table~\ref{tab:phfu}). After detecting the signal we re-applied the TFA
filter, this time in signal reconstruction mode, so as to obtain an
undistorted trend-filtered light curve. The per--point root mean square (RMS)
residual combined filtered HATSouth light curve (after subtracting the
best-fit model transit) is 0.015\,mag, which is typical for a star of this
magnitude.

\subsubsection{Photometric follow-up}
\label{sec:phfu}

\begin{figure}[!ht]
\plotone{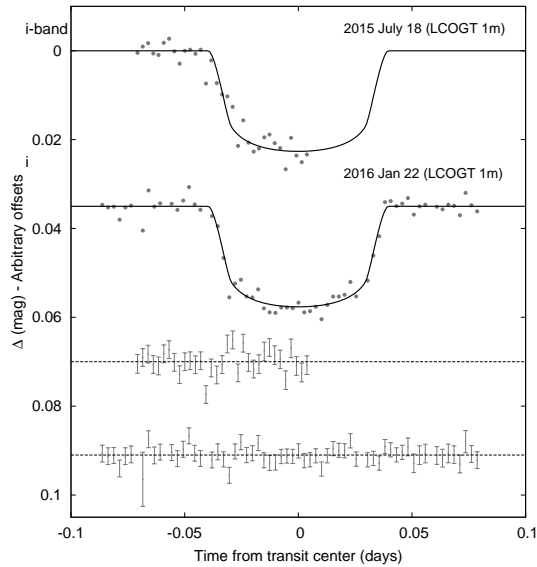}
\caption{
        Unbinned follow-up transit light curve of \hatcur{} obtained with
        telescopes from the LCOGT~1\,m network. Our best fit is shown by the
        solid lines. The residuals from the best-fit model are shown below in
        the same order.\\
\label{fig:lc}} \end{figure}

We obtained follow-up light curves of \hatcur{} using the LCOGT~1\,m
telescope network. An ingress was observed on UT 2015 July 18 with the SBIG
camera and a Sloan $i$ filter on the 1\,m at the South African Astronomical
Observatory (SAAO). A total of 33 images were collected at a median cadence
of 201\,s. A full transit was observed on UT 2016 Jan 22 with the sinistro
camera and a Sloan $i$ filter on the 1\,m at Cerro Tololo Inter-American
Observatory (CTIO). A total of 61 images were collected at a median cadence
of 219\,s. For the record we also note that a full transit was observed on UT
2016 January 3 with the SBIG camera on the 1\,m at SAAO, however due to
tracking and weather problems we were unable to extract high precision
photometry from these images, and therefore do not include these data in our
analysis. For details of the reduction procedure used to extract light curves
from the raw images see \citet{penev:2013:hats1}. The follow-up light curves
are shown, together with our best-fit model, in Figure~\ref{fig:lc}, while
the data are available in Table~\ref{tab:phfu}. The per-point precision of
the SBIG observations is 2.5\,mmag, while the per-point precision of the
sinistro observations is 1.7\,mmag.

\ifthenelse{\boolean{emulateapj}}{
        \begin{deluxetable*}{lrrrrr} }{
        \begin{deluxetable}{lrrrrr}
    }
        \tablewidth{0pc}
        \tablecaption{Differential photometry of \hatcur\label{tab:phfu}}
        \tablehead{ \colhead{BJD} &
        \colhead{Mag\tablenotemark{a}} &
        \colhead{\ensuremath{\sigma_{\rm Mag}}} &
        \colhead{Mag(orig)\tablenotemark{b}} & \colhead{Filter} &
        \colhead{Instrument} \\ \colhead{\hbox{~~~~(2\,400\,000$+$)~~~~}} &
        \colhead{} & \colhead{} & \colhead{} & \colhead{} & \colhead{} }
        \startdata
            $ 56442.67216 $ & $   0.03291 $ & $   0.00789 $ & $ \cdots $ & $ r$ &         HS\\
$ 56411.67208 $ & $  -0.01713 $ & $   0.00723 $ & $ \cdots $ & $ r$ &         HS\\
$ 56343.80691 $ & $  -0.00841 $ & $   0.00655 $ & $ \cdots $ & $ r$ &         HS\\
$ 56444.34817 $ & $   0.01720 $ & $   0.00712 $ & $ \cdots $ & $ r$ &         HS\\
$ 56392.40213 $ & $  -0.00247 $ & $   0.00646 $ & $ \cdots $ & $ r$ &         HS\\
$ 56395.75361 $ & $   0.02231 $ & $   0.00737 $ & $ \cdots $ & $ r$ &         HS\\
$ 56416.69970 $ & $  -0.02166 $ & $   0.00680 $ & $ \cdots $ & $ r$ &         HS\\
$ 56469.48392 $ & $  -0.02100 $ & $   0.00717 $ & $ \cdots $ & $ r$ &         HS\\
$ 56446.86219 $ & $  -0.00054 $ & $   0.00641 $ & $ \cdots $ & $ r$ &         HS\\
$ 56458.59202 $ & $  -0.03131 $ & $   0.00744 $ & $ \cdots $ & $ r$ &         HS\\

            [-1.5ex]
\enddata \tablenotetext{a}{
     The out-of-transit level has been subtracted. For the HATSouth light
     curve (rows with ``HS'' in the Instrument column), these magnitudes have
     been detrended using the EPD and TFA procedures prior to fitting a
     transit model to the light curve. We apply the TFA in
     signal-reconstruction mode so as to preserve the transit depth. For the
     follow-up light curves (rows with an Instrument other than ``HS'') these
     magnitudes have been detrended with the EPD procedure, carried out
     simultaneously with the transit fit.
}
\tablenotetext{b}{
        Raw magnitude values without application of the EPD procedure.  This
        is only reported for the follow-up light curves.
}
\tablecomments{
        This table is available in a machine-readable form in the online
        journal.  A portion is shown here for guidance regarding its form and
        content. The data are also available on the HATSouth website at
        \url{http://www.hatsouth.org}.
} \ifthenelse{\boolean{emulateapj}}{ \end{deluxetable*} }{ \end{deluxetable} }

\subsection{Spectroscopy}
\label{sec:hispec}

Spectroscopic follow-up observations of \hatcur{} were carried out with WiFeS
on the ANU~2.3\,m telescope \citep{dopita:2007} and with FEROS on the
MPG~2.2\,m \citep{kaufer:1998}.

A total of three spectra were obtained with WiFeS between UT 2015 Feb 28 and
UT 2015 Mar 2, two at a resolution of $R \equiv \Delta\,\lambda\,/\,\lambda =
7000$, and one at $R = 3000$. These data were reduced and analyzed following
the procedure described by \citet{bayliss:2013:hats3}. The $R = 3000$
spectrum was used to estimate the spectral type and surface gravity of
\hatcur{} (we find that it is a G dwarf), while the $R = 7000$ spectra were
used to rule out an RV variation greater than 5\,\kms. 

\begin{figure} [ht]
\plotone{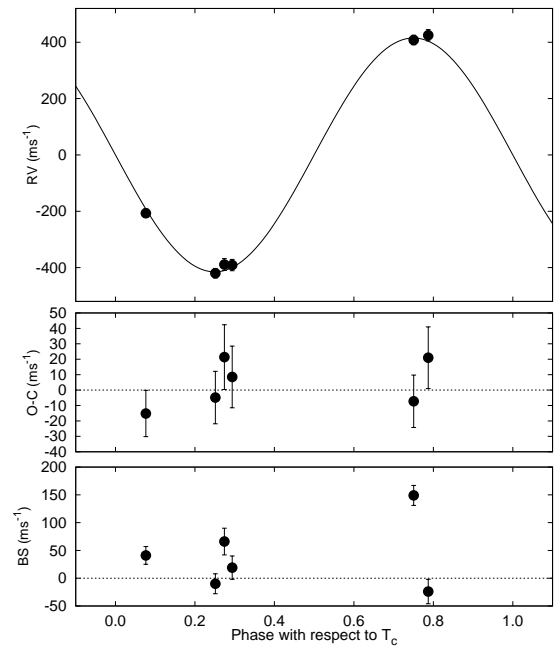}
\caption{
    {\em Top panel:} High-precision RV measurements from MPG~2.2\,m/FEROS
    together with our best-fit orbit model.  Zero phase corresponds to the
    time of mid-transit.  The center-of-mass velocity has been subtracted.
    {\em Second panel:} Velocity $O\!-\!C$ residuals from the best-fit model.
    The error bars include the jitter which is varied in the fit.  {\em Third
    panel:} Bisector spans (BS).  Note the different vertical scales of the
    panels.\\
\label{fig:rvbis}} \end{figure}

We obtained six $R = 48000$ spectra with FEROS between UT 2015 Jun 12 and UT
2015 Jun 20. These were reduced to high precision RV and spectral line
bisector span (BS) measurements following \citet{jordan:2014:hats4}, and were
also used to determine high precision atmospheric parameters
(Section~\ref{sec:analysis}). The RVs show a clear $K = \hatcurRVK{}$\,\ms\
sinusoidal variation in phase with the transit ephemeris (\reffigl{rvbis};
the data are provided in \reftabl{rvs}), confirming this object as a
transiting planet system. The BSs exhibit significant scatter, as is typical
for a faint $V = \hatcurCCtassmv$\,mag star, but are uncorrelated with the
RVs. The scatter is also well below the level expected if this were a blended
stellar eclipsing binary system (Section~\ref{sec:analysis}).

%

\ifthenelse{\boolean{emulateapj}}{ \begin{deluxetable*}{lrrrrrr} }{
    \begin{deluxetable}{lrrrrrr} } \tablewidth{0pc} \tablecaption{
    Relative radial velocities and bisector span measurements of \hatcur{}.
    \label{tab:rvs}
} \tablehead{ \colhead{BJD} & \colhead{RV\tablenotemark{a}} &
\colhead{\ensuremath{\sigma_{\rm RV}}\tablenotemark{b}} & \colhead{BS} &
\colhead{\ensuremath{\sigma_{\rm BS}}} & 
        \colhead{Phase} & \colhead{Instrument}\\
    \colhead{\hbox{(2\,457\,100$+$)}} & \colhead{(\ms)} & \colhead{(\ms)} &
    \colhead{(\ms)} &
    \colhead{} &
        \colhead{} & \colhead{} } \startdata $ 85.64999 $ & $  -389.04 $ & $    21.00 $ & $   66.0 $ & $   24.0 $ & $   0.274 $ & FEROS \\
$ 86.50430 $ & $  -391.04 $ & $    20.00 $ & $   19.0 $ & $   21.0 $ & $   0.294 $ & FEROS \\
$ 88.59324 $ & $   424.96 $ & $    20.00 $ & $  -24.0 $ & $   22.0 $ & $   0.787 $ & FEROS \\
$ 90.51136 $ & $  -207.04 $ & $    15.00 $ & $   41.0 $ & $   16.0 $ & $   0.076 $ & FEROS \\
$ 91.49572 $ & $  -420.04 $ & $    17.00 $ & $  -10.0 $ & $   18.0 $ & $   0.251 $ & FEROS \\
$ 93.58965 $ & $   407.96 $ & $    17.00 $ & $  149.0 $ & $   18.0 $ & $   0.750 $ & FEROS \\

        [-1.5ex] \enddata \tablenotetext{a}{
    Relative RVs, with $\gamma_{RV}$ (see table~\ref{tab:stellar})
    subtracted.
} \tablenotetext{b}{
        Internal errors excluding the component of astrophysical/instrumental
        jitter considered in \refsecl{analysis}.
}
\ifthenelse{\boolean{emulateapj}}{ \end{deluxetable*} }{ \end{deluxetable} }

\section{Analysis}
\label{sec:analysis}

\begin{figure}[]
\plotone{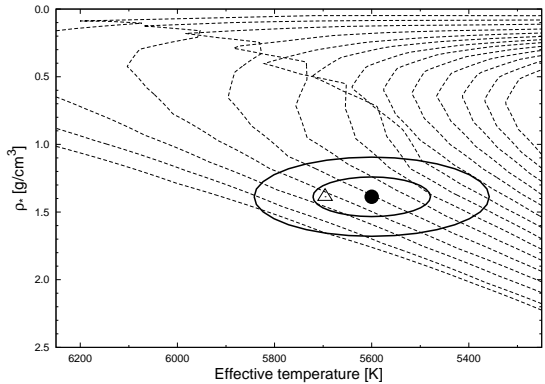}
\caption[]{
    Comparison between the measured values of \teffstar\ and \rhostar\ (from
    ZASPE applied to the FEROS spectra, and from our modeling of the light
    curves and RV data, respectively), and the Y$^{2}$ model isochrones from
    \citet{yi:2001}. The best-fit values (dark filled circle), and
    approximate 1$\sigma$ and 2$\sigma$ confidence ellipsoids are shown. The
    values from our initial ZASPE iteration are shown with the open triangle.
    The Y$^{2}$ isochrones are shown for ages of 0.2\,Gyr, and 1.0 to
    14.0\,Gyr in 1\,Gyr increments.
\label{fig:iso}} \end{figure}

We analyzed the photometric and spectroscopic observations of \hatcur{} to
determine the parameters of the system using the standard procedures
developed for HATNet and HATSouth (see \citealp{bakos:2010:hat11}, with
modifications described by \citealp{hartman:2012:hat39hat41}).

High-precision stellar atmospheric parameters were measured from the FEROS
spectra using ZASPE \citep{brahm:2016:zaspe}. The resulting \teffstar\ and
\feh\ measurements were combined with the stellar density \rhostar\
determined through our joint light curve and RV curve analysis, to determine
the stellar mass, radius, age, luminosity, and other physical parameters, by
comparison with the Yonsei-Yale \citep[Y$^{2}$;][]{yi:2001} stellar evolution
models (see Figure~\ref{fig:iso}). This provided a revised estimate of
\loggstar\ which was fixed in a second iteration of ZASPE. Our final adopted
stellar parameters are listed in Table~\ref{tab:stellar}. We find that the
star \hatcur{} has a mass of \hatcurISOm\,\msun, a radius of
\hatcurISOr\,\rsun, and is at a reddening-corrected distance of
\hatcurXdistred\,pc.

We simultaneously carried out a joint analysis of the High-precision FEROS
RVs (fit using a Keplerian orbit) and the HS and LCOGT~1\,m light curves (fit
using a \citealp{mandel:2002} transit model with fixed quadratic limb
darkening coefficients taken from \citealp{claret:2004}) to measure the
stellar density, as well as the orbital and planetary parameters. This
analysis makes use of a differential evolution Markov Chain Monte Carlo
procedure \citep[DEMCMC;][]{terbraak:2006} to estimate the posterior
parameter distributions, which we use to determine the median parameter
values and their 1$\sigma$ uncertainties. The results are listed in
Table~\ref{tab:planetparam}. We find that the planet \hatcurb{} has a mass of
\hatcurPPmlong\,\mjup, and a radius of \hatcurPPrlong\,\rjup. We fit the data
both assuming a circular orbit, and allowing for a non-zero eccentricity. We
find that the observations are consistent with a circular orbit: $e =
\hatcurRVecceneccen$, with a 95\% confidence upper-limit of
$e\hatcurRVeccentwosiglimeccen{}$, and therefore adopt the parameters that
come from assuming a circular orbit (we also find that the Bayesian evidence
for the circular orbit model is higher than the evidence for the
free-eccentricity model).

\subsection{Ruling Out Blended Models}
In order to rule out the possibility that \hatcur{} is a blended
stellar eclipsing binary system, we carried out a blend analysis of
the photometric data following \citet{hartman:2012:hat39hat41}. We
find that all blend models tested can be rejected based on the
photometry alone with $3.5\sigma$ confidence. Moreover, the blend
models which come closest to fitting the photometry (those which
cannot be rejected with greater than $5\sigma$ confidence) yield
simulated RVs that are not at all similar to what we observe (i.e.,
the simulated blend-model RVs do not show a sinusoidal variation in
phase with the photometric ephemeris). We conclude that \hatcur{} is
not a blended stellar eclipsing binary system, and is instead a
transiting planet system.

\subsection{Photometric Rotation Period}
\label{sec:rotation}

\begin{figure}[]
    \plotone{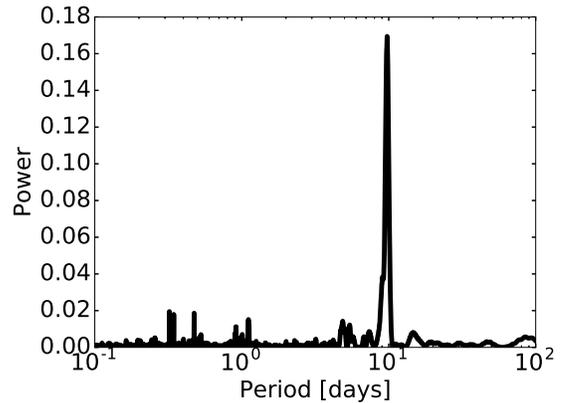}
    \plotone{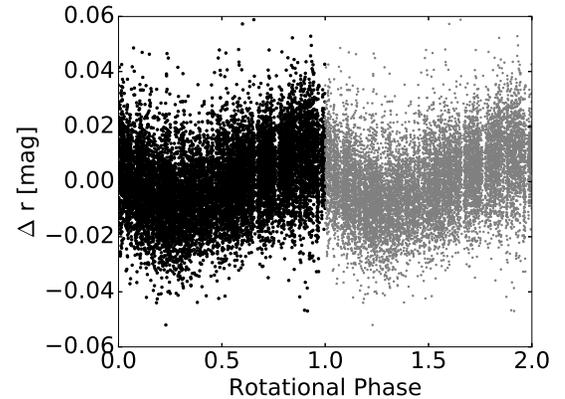}
    \caption[]{
        Top: The Lomb--Scargle periodogram of \hatcur{} light curves (in signal
            reconstruction mode for the transits but not the rotational
            modulation) with transits removed. Bottom: the same lightcurve
            folded with the best--fit stellar spin period (the points in the
            second half of the plot are duplicates of those in the first half).
    }
    \label{fig:hats18_spin}
\end{figure}

The lightcurve of \hatcur{} shows a clear signature of stellar spin
variability. In Fig.\,\ref{fig:hats18_spin} we show the Lomb--Scargle
periodogram of the HATSouth discovery lightcurve of HATS-18, with observations
during transits removed, as well as the lightcurve as a function of the best
fit spin period (9.8 days) phase. In order to get a handle on the uncertainty
in the stellar spin period we split the lightcurve into 9 segments, each
containing three spin periods and fit for the rotation period in each segment
separately and adopt the standard deviation of the individual measurements as
the period uncertainty. The resulting spin period estimate is $P_{rot\star} =
9.8 \pm 0.4$ days.

\ifthenelse{\boolean{emulateapj}}{
  \begin{deluxetable*}{lcr}
}{
  \begin{deluxetable}{lcr}
}
\tablewidth{0pc}
\tabletypesize{\scriptsize}
\tablecaption{
    Stellar Parameters for \hatcur{} 
    \label{tab:stellar}
}
\tablehead{
    \multicolumn{1}{c}{~~~~~~~~Parameter~~~~~~~~} &
    \multicolumn{1}{c}{Value}                     &
    \multicolumn{1}{c}{Source}    
}
\startdata
\noalign{\vskip -3pt}
\sidehead{Identifying Information}
~~~~R.A.~(h:m:s)                      &  \hatcurCCra{} & 2MASS\\
~~~~Dec.~(d:m:s)                      &  \hatcurCCdec{} & 2MASS\\
~~~~R.A.p.m.~(mas/yr)                 &  \hatcurCCpmra{} & 2MASS\\
~~~~Dec.p.m.~(mas/yr)                 &  \hatcurCCpmdec{} & 2MASS\\
~~~~GSC ID                            &  \hatcurCCgsc{} & GSC\\
~~~~2MASS ID                          &  \hatcurCCtwomass{} & 2MASS\\
\sidehead{Spectroscopic properties}
~~~~$\teffstar$ (K)\dotfill         &  \hatcurSMEteff{} & ZASPE \tablenotemark{a}\\
~~~~Spectral type\dotfill           &  \hatcurISOspec & ZASPE\\
~~~~$\feh$\dotfill                  &  \hatcurSMEzfeh{} & ZASPE                 \\
~~~~$\vsini$ (\kms)\dotfill         &  \hatcurSMEvsin{} & ZASPE                 \\
~~~~$\gamma_{\rm RV}$ (\ms)\dotfill&  \hatcurRVgammaabs{} & FEROS                  \\
\sidehead{Photometric properties}
~~~~$B$ (mag)\dotfill               &  \hatcurCCtassmB{} & APASS                \\
~~~~$V$ (mag)\dotfill               &  \hatcurCCtassmv{} & APASS               \\
~~~~$g$ (mag)\dotfill               &  \hatcurCCtassmg{} & APASS                \\
~~~~$r$ (mag)\dotfill               &  \hatcurCCtassmr{} & APASS                \\
~~~~$i$ (mag)\dotfill               &  \hatcurCCtassmi{} & APASS                \\
~~~~$J$ (mag)\dotfill               &  \hatcurCCtwomassJmag{} & 2MASS           \\
~~~~$H$ (mag)\dotfill               &  \hatcurCCtwomassHmag{} & 2MASS           \\
~~~~$K_s$ (mag)\dotfill             &  \hatcurCCtwomassKmag{} & 2MASS           \\
\sidehead{Derived properties}
~~~~$\mstar$ ($\msun$)\dotfill      &  \hatcurISOmlong{} & Y$^{2}$+\hatcurlumind{}+ZASPE \tablenotemark{b}\\
~~~~$\rstar$ ($\rsun$)\dotfill      &  \hatcurISOrlong{} & Y$^{2}$+\hatcurlumind{}+ZASPE         \\
~~~~$\loggstar$ (cgs)\dotfill       &  \hatcurISOlogg{} & Y$^{2}$+\hatcurlumind{}+ZASPE         \\
~~~~$\rhostar$ (\gcmc) \tablenotemark{c}\dotfill & \hatcurLCrho{} & Light curves \\
~~~~$\rhostar$ (\gcmc) \tablenotemark{c}\dotfill & \hatcurISOrho{} & Y$^{2}$+Light curves+ZASPE \\
~~~~$\lstar$ ($\lsun$)\dotfill      &  \hatcurISOlum{} & Y$^{2}$+\hatcurlumind{}+ZASPE         \\
~~~~$M_V$ (mag)\dotfill             &  \hatcurISOmv{} & Y$^{2}$+\hatcurlumind{}+ZASPE         \\
~~~~$M_K$ (mag,\hatcurjhkfilset{})&  \hatcurISOMK{} & Y$^{2}$+\hatcurlumind{}+ZASPE         \\
~~~~Age (Gyr)\dotfill               &  \hatcurISOage{} & Y$^{2}$+\hatcurlumind{}+ZASPE         \\
~~~~$A_{V}$ (mag) \tablenotemark{d}\dotfill           &  \hatcurXAv{} & Y$^{2}$+\hatcurlumind{}+ZASPE\\
~~~~Distance (pc)\dotfill           &  \hatcurXdistred{} & Y$^{2}$+\hatcurlumind{}+ZASPE\\
~~~~$P_{rot\star}$ (days)\dotfill   &  $9.8\pm0.4$ & HATSouth light curve\\
\enddata
\tablenotetext{a}{
    ZASPE = ``Zonal Atmospherical Stellar Parameter Estimator'' method
    for the analysis of high-resolution spectra
    applied to the FEROS spectra of \hatcur{}. These parameters rely
    primarily on ZASPE, but have a small dependence also on the
    iterative analysis incorporating the isochrone search and global
    modeling of the data, as described in the text.  }
\tablenotetext{b}{
    Isochrones+\hatcurlumind{}+ZASPE = Based on the Y$^{2}$ isochrones
    \citep{yi:2001},
    the stellar density used as a luminosity indicator, and the ZASPE
    results.
} 
\tablenotetext{c}{We list two values for $\rhostar$. The first value is
determined from the global fit to the light curves and RV data, without
imposing a constraint that the parameters match the stellar evolution models. The second value results from restricting the posterior distribution to combinations of $\rhostar$+$\teffstar$+$\feh$ that match to a Y$^{2}$ stellar model.}
\tablenotetext{d}{ Total \band{V} extinction to the star determined
  by comparing the catalog broad-band photometry listed in the table
  to the expected magnitudes from the
  Isochrones+\hatcurlumind{}+ZASPE model for the star. We use the
  \citet{cardelli:1989} extinction law.  }
\ifthenelse{\boolean{emulateapj}}{
  \end{deluxetable*}
}{
  \end{deluxetable}
}

\ifthenelse{\boolean{emulateapj}}{
  \begin{deluxetable*}{lr}
}{
  \begin{deluxetable}{lr}
}
\tabletypesize{\scriptsize}
\tablecaption{Parameters for the transiting planet \hatcurb{}.
\label{tab:planetparam}}
\tablehead{
    \multicolumn{1}{c}{~~~~~~~~Parameter~~~~~~~~} &
    \multicolumn{1}{r}{Value \tablenotemark{a}}                     
}
\startdata
\noalign{\vskip -3pt}
\sidehead{\Lc{} parameters}
~~~$P$ (days)             \dotfill    & $\hatcurLCP{}$              \\
~~~$T_c$ (${\rm BJD}$)    
      \tablenotemark{b}   \dotfill    & $\hatcurLCT{}$              \\
~~~$T_{14}$ (days)
      \tablenotemark{b}   \dotfill    & $\hatcurLCdur{}$            \\
~~~$T_{12} = T_{34}$ (days)
      \tablenotemark{b}   \dotfill    & $\hatcurLCingdur{}$         \\
~~~$\arstar$              \dotfill    & $\hatcurPPar{}$             \\
~~~$\zrstar$ \tablenotemark{c}              \dotfill    & $\hatcurLCzeta{}$\phn       \\
~~~$\rpl/\rstar$          \dotfill    & $\hatcurLCrprstar{}$        \\
~~~$b^2$                  \dotfill    & $\hatcurLCbsq{}$            \\
~~~$b \equiv a \cos i/\rstar$
                          \dotfill    & $\hatcurLCimp{}$           \\
~~~$i$ (deg)              \dotfill    & $\hatcurPPi{}$\phn         \\

\sidehead{Limb-darkening coefficients \tablenotemark{d}}
~~~$c_1,i$ (linear term)  \dotfill    & $\hatcurLBii{}$            \\
~~~$c_2,i$ (quadratic term) \dotfill  & $\hatcurLBiii{}$           \\
~~~$c_1,r$               \dotfill    & $\hatcurLBir{}$             \\
~~~$c_2,r$               \dotfill    & $\hatcurLBiir{}$            \\

\sidehead{RV parameters}
~~~$K$ (\ms)              \dotfill    & $\hatcurRVK{}$\phn\phn      \\
~~~$e$ \tablenotemark{e}  \dotfill    & $\hatcurRVeccentwosiglimeccen{}$ \\
~~~FEROS RV jitter (\ms) \tablenotemark{f}        \dotfill    & \hatcurRVjittertwosiglim{}           \\

\sidehead{Planetary parameters}
~~~$\mpl$ ($\mjup$)       \dotfill    & $\hatcurPPmlong{}$          \\
~~~$\rpl$ ($\rjup$)       \dotfill    & $\hatcurPPrlong{}$          \\
~~~$C(\mpl,\rpl)$
    \tablenotemark{g}     \dotfill    & $\hatcurPPmrcorr{}$         \\
~~~$\rhopl$ (\gcmc)       \dotfill    & $\hatcurPPrho{}$            \\
~~~$\log g_p$ (cgs)       \dotfill    & $\hatcurPPlogg{}$           \\
~~~$a$ (AU)               \dotfill    & $\hatcurPParel{}$          \\
~~~$T_{\rm eq}$ (K) \tablenotemark{h}        \dotfill   & $\hatcurPPteff{}$           \\
~~~$\Theta$ \tablenotemark{i} \dotfill & $\hatcurPPtheta{}$         \\
~~~$\langle F \rangle$ ($10^{\hatcurPPfluxavgdim}$\ergscmsq) \tablenotemark{i}
                          \dotfill    & $\hatcurPPfluxavg{}$       \\ [-1.5ex]
\enddata
\tablenotetext{a}{
    For each parameter we give the median value and
    68.3\% (1$\sigma$) confidence intervals from the posterior
    distribution. Reported results assume a circular orbit.
}
\tablenotetext{b}{
    Reported times are in Barycentric Julian Date calculated directly
    from UTC, {\em without} correction for leap seconds.
    \ensuremath{T_c}: Reference epoch of mid transit that
    minimizes the correlation with the orbital period.
    \ensuremath{T_{14}}: total transit duration, time
    between first to last contact;
    \ensuremath{T_{12}=T_{34}}: ingress/egress time, time between first
    and second, or third and fourth contact.
}
\tablenotetext{c}{
    Reciprocal of the half duration of the transit used as a jump
    parameter in our MCMC analysis in place of $\arstar$. It is
    related to $\arstar$ by the expression $\zrstar = \arstar
    (2\pi(1+e\sin \omega))/(P \sqrt{1 - b^{2}}\sqrt{1-e^{2}})$
    \citep{bakos:2010:hat11}.
}
\tablenotetext{d}{
    Values for a quadratic law, adopted from the tabulations by
    \cite{claret:2004} according to the spectroscopic (ZASPE) parameters
    listed in \reftabl{stellar}.
}
\tablenotetext{e}{
    The 95\% confidence upper-limit on the eccentricity. All other
    parameters listed are determined assuming a circular orbit.
}
\tablenotetext{f}{
    Error term, either astrophysical or instrumental in origin, added
    in quadrature to the formal RV errors. This term is varied in the
    fit assuming a prior that is inversely proportional to the jitter. We find that the jitter is consistent with zero, and thus give the 95\% confidence upper limit.
}
\tablenotetext{g}{
    Correlation coefficient between the planetary mass \mpl\ and
    radius \rpl\ determined from the parameter posterior distribution
    via $C(\mpl,\rpl) = \langle(\mpl - \langle\mpl\rangle)(\rpl -
    \langle\rpl\rangle)\rangle/(\sigma_{\mpl}\sigma_{\rpl})\rangle$, 
	where $\langle \cdot \rangle$ is the
    expectation value operator, and $\sigma_x$ is the standard
    deviation of parameter $x$.
}
\tablenotetext{h}{
    Planet equilibrium temperature averaged over the orbit, calculated
    assuming a Bond albedo of zero, and that flux is re--radiated from
    the full planet surface.
}
\tablenotetext{i}{
    The Safronov number is given by $\Theta = \frac{1}{2}(V_{\rm
    esc}/V_{\rm orb})^2 = (a/\rpl)(\mpl / \mstar )$
    \citep[see][]{hansen:2007}.
}
\tablenotetext{j}{
    Incoming flux per unit surface area, averaged over the orbit.
}
\ifthenelse{\boolean{emulateapj}}{
  \end{deluxetable*}
}{
  \end{deluxetable}
}
%


\section{Comparison to Other Short Period Systems}
\label{sec:comparison}

\begin{figure}[]
\plotone{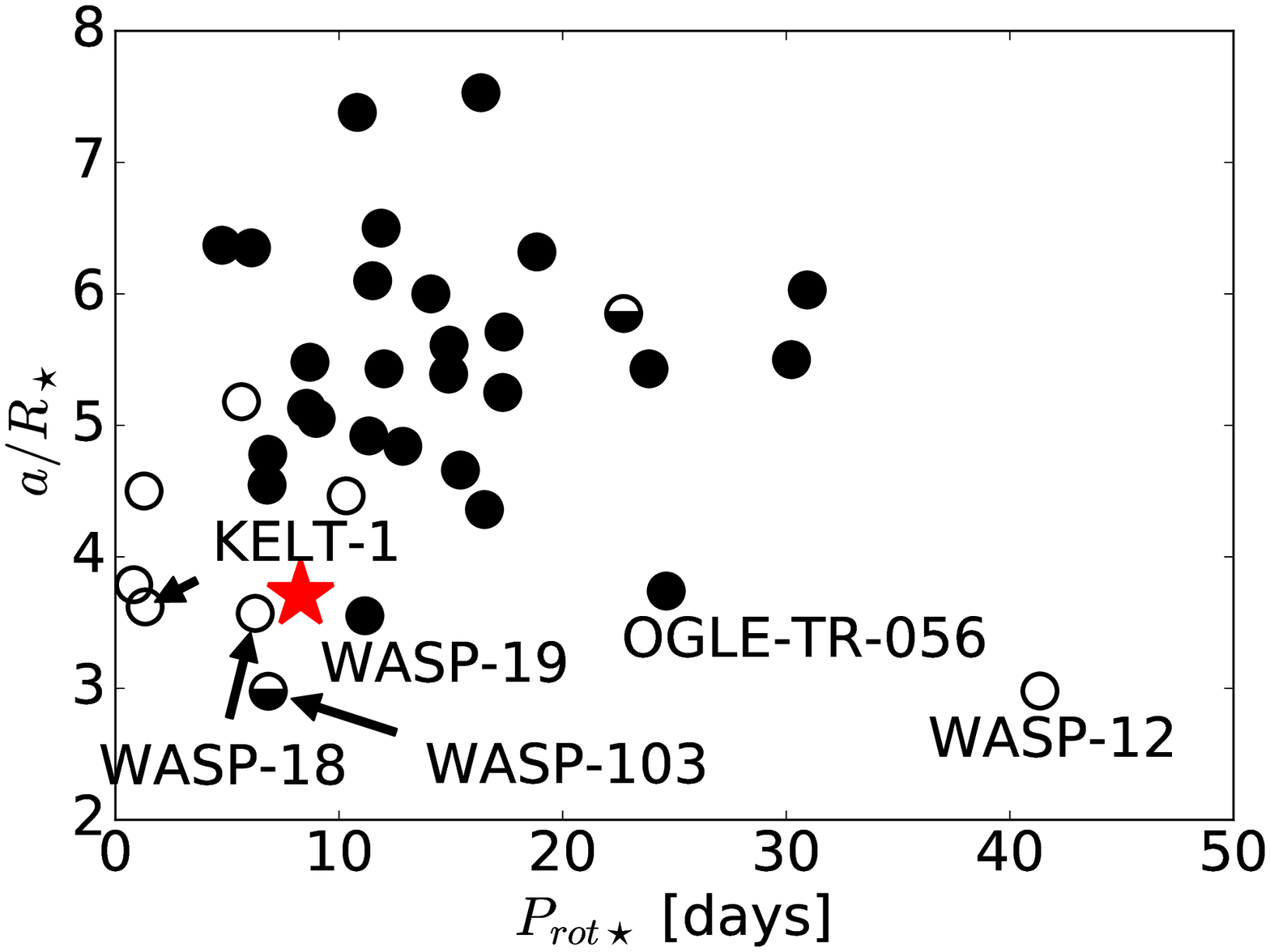}
\caption[]{
        Size of the planetary orbit relative to the stellar radius as a
        function of the stellar rotation period, estimated using the measured
        projected equatorial velocity of the stars and their estimated radii.
        Planets other than \hatcur{}(big star symbol) are all transiting
        planets from the NASA Exoplanet Archive with orbital periods shorter
        than 2 days and masses at least $0.1$\,\mjup. Filled symbols: host
        star effective temperature is below 6250\,K (surface convective zone
        stars); empty symbols: host star effective temperature is above
        6250\,K (surface radiative zone stars); half--filled symbols: host
        star effective temperature is consistent with 6250\,K within quoted
        error bars.
} \label{fig:aR_spin}
\label{fig:hats18_first_comparison}
\end{figure}

\begin{figure}[]
\plotone{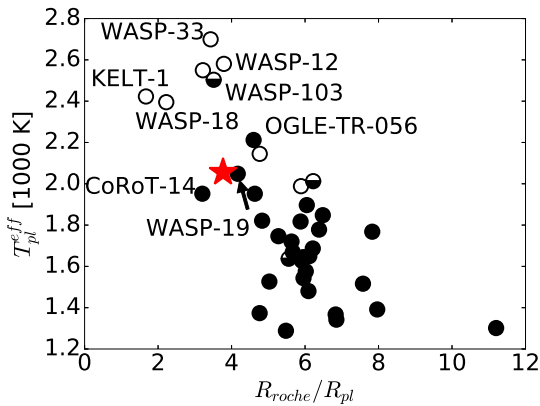}
\caption[]{
        The equilibrium temperature of the planet, assuming ideal black body
        against the fraction of the ratio of the Roche radius to planet radius
        for the same systems plotted in
        Fig.\,\ref{fig:hats18_first_comparison}.
} \label{fig:Teff_roche}
\end{figure}

\begin{figure}[]
\plotone{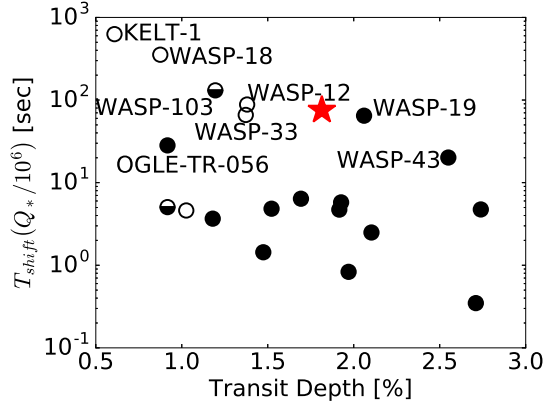}
\caption[]{
        The shift in mid--transit time ephemeris after a decade for a tidal
        quality factor of $Q_\star = 10^6$.
} \label{fig:tshift_depth} \label{fig:hats18_last_comparison}.
\end{figure}

Due to its very short orbital period and relatively high planetary mass,
the HATS-18 system is ideal for testing theories of star--planet
interactions, whether those occur through radiation, gravity or magnetic
fields. Figures
\ref{fig:hats18_first_comparison}---\ref{fig:hats18_last_comparison} show a
comparison between the present sample of giant planets (mass at least
$0.1$\,\mjup) in orbital periods shorter than two days and the \hatcur{}
system in a number of parameters related to the strength of various
star--planet interactions that have been suggested to occur.

The possible magnetic interactions (and hence their observable effects) are
expected to grow in strength the deeper the planet is in its star's magnetic
field and the stronger the field is. In general, stars with surface
convective zones are expected to have much stronger magnetic fields than
stars with surface radiative zones, since in the former case some form of
convectively driven dynamo is expected to operate in the stellar envelope.
Further, the dynamo is expected to generate a larger field for faster
rotating stars, hence the two readily observable quantities to compare in
order to gauge the observability of magnetic star--planet interactions are
the size of the orbit relative to the stellar radius (\arstar{}) and the
stellar spin period. From Fig.\,\ref{fig:aR_spin}, we see that \hatcur{} is
among the three surface convective zone systems (\hatcur{}, WASP-19 and
OGLE-TR-56) whose error bars are consistent with having the smallest
\arstar{}\, and among those it has the shortest stellar spin period (inferred
either from its projected spin velocity, or the observed rotational
modulation in its lightcurve).

Another rather dramatic effect of star--planet interactions is for the stellar
irradiation/wind to drive outflows from the planet. Clearly this process will
occur more readily for planets closer to filling their Roche radius and for
hotter planets. Fig.\,\ref{fig:Teff_roche} plots the ratio of the planetary to
the Roche radius for each system against the equilibrium effective temperature
for the planet (assuming a perfect black body) for the same sample of planets
as in Fig.\,\ref{fig:hats18_first_comparison}. Again, \hatcur{} is among the
planets with most favourable parameters, although in this case there is a
cluster of very--hot, very small Roche ratio planets around surface radiative
zone stars, for one of which (WASP-12b) outflows have been claimed
\citep[c.f.][]{Fossati_et_al_10, Haswell_et_al_12}.

The most direct way of detecting tidal interactions between a star and its
companion planet is to see the orbital decay due to tidal dissipation in the
star. This is most readily accomplished through observing the resulting
deviation from a linear mid--transit time ephemeris. Detecting this effect will
provide a direct measurement of the tidal dissipation efficiency of the parent
star: the least constrained parameter in tidal interactions involving stars and
giant planets. Fig.\,\ref{fig:tshift_depth} shows that \hatcurb{} is the planet
around a convective envelope star with the largest expected shift in
mid--transit time after a decade.

\section{Host Star Spin--Up and a measurement of $Q_\star$}
\label{sec:Q}
Given that \hatcur{} has an age consistent with the age of the Sun, and that
it is very close to solar mass, its spin period should be close to that of
the Sun or to the recently measured rotation periods in the 4.2\,Gyr old open
cluster M\,67 \citep{Barnes_et_al_16}: $P_{rot\star}\approx30$ days, even if
the stellar age were at the lower end of the estimated error bar (2.2 Gyrs),
the expected spin period is $P_{rot\star}\approx20$. Instead, in
\refsec{analysis} we found $\vsini$ and stellar radius corresponding to a
spin period of $P_{rot\star} = 8.3 \pm 0.8$ days, which is consistent with
the photometrically determined rotation period of $P_{rot\star} = 9.8 \pm
0.4$ days. This much faster spin rate is close to what is observed for Solar
mass stars in clusters with ages around 600\,Myr: the 550\,Myr old M\,37
\citep{Hartman_et_al_09}, the 580\,Myr old Praesepe \citep{Agueros_et_al_11,
Delorme_et_al_11,Kovacs_et_al_14}, and the 625\,Myr old Hyades
\citep{Delorme_et_al_11}. A natural explanation for this apparent discrepancy
is suggested by the fact that the \hatcur{} system contains a very short
period giant planet, which should have experienced some orbital decay due to
tidal dissipation in the star. The angular momentum taken out of the
planetary orbit as it shrinks is deposited in the star and hence the star is
spun--up.  The fact that we see evidence for this tidal spin--up, means that
we can use it to measure the tidal dissipation properties of the star. In
this section we describe a method for carrying out such a measurement and
show the resulting constraints.

\subsection{The Tidal and Stellar Spin Model}
\label{sec:Q_model}
Stars like \hatcur{} continuously lose angular momentum throughout their
lifetime by magnetically imparting angular momentum to the wind of charged
particles launched from their surfaces. As a result, in order to relate the
stellar tidal dissipation efficiency to the observed stellar spin, we need to
model this angular momentum loss simultaneously with the tidal spin--up.

There are a number of options for modelling the tidal evolution, and the
angular momentum loss. However, in an effort to keep the number of model
parameters small while constructing a consistent model we will use the tidal
evolution formulation of \citep{Lai_12} and assume a constant value for
$Q'_\star \equiv Q_\star/k_2$, where $Q_\star$ is the fraction of tidal
energy lost in one orbital period, and $k_2$ is the Love number of the
star.  Note that, while tidal dissipation in the planet may be more
efficient than in the star, it will quickly result in a circular orbit and
planetary spin synchronized with the orbit, which will make the tidal
deformation of the planet static and hence not subject to dissipation.
Further, assuming constant dissipation efficiency is clearly not physical.
In particular, the dissipation should vanish ($Q'_\star = \infty$) when the
tidal frequency approaches zero and increase gradually as the frequency
moves away from zero. However, for tidal frequencies near that observed for
\hatcur{}, the dissipation is expected to become less efficient as the
frequency increases.  Since there is currently no agreement on the expected
dependence of $Q'_\star$ on frequency and other parameters, we don't have a
choice but to assume $Q'_\star = const$. In practice, the way to interpret
the results is that the $Q'_\star$ measured by our analysis is appropriate
for the currently observed state of the system analyzed, since the observed
spin--up of the host star is overwhelmingly dominated by the very recent
tidal evolution (see Fig. \ref{fig:sample_evolutions}). 

We will model the star as consisting of two distinct zones: the surface
convective envelope and the radiative core, and all tidal dissipation will be
assumed to occur in the envelope. As a result, any angular momentum lost by
the orbit will be deposited exclusively in the convective zone of \hatcur{}.
This will tend to drive differential rotation between the core and the
envelope, which will in turn be suppressed by at present not well understood
coupling processes, but its efficiency is reasonably constrained by
observations \citep[c.f.][]{Irwin_et_al_07, Gallet_Bouvier_15,
Amard_et_al_16}. The model for the evolution of the stellar spin tracks a
single value for the spin of each zone, allows for angular momentum exchange
between the core and the envelope and for angular momentum loss due to the
stellar wind. The particular formulation we will use is given in detail in
\citet{Irwin_et_al_07}. 

The loss of angular momentum from the convective envelope due to the wind is
given by:
\begin{equation}
    \begin{split}
        \left(\frac{d \vec{J}_{conv}}{d t}\right)_{wind}
        \equiv&
        \quad K \vec{\omega}_{conv} \min(\left|\vec{\omega}_{conv}\right|^2,
        \omega_{sat}^2) \\
        & \left(\frac{R_\star}{R_\odot}\right)^{1/2}
        \left(\frac{M_\star}{M_\odot}\right)^{-1/2}
    \end{split}
\end{equation}
Where $K$ and $\omega_{sat}$ are parameters for the efficiency of the
coupling of the convective zone rotation to the wind, $\vec{J}_{conv}$ is the
angular momentum of the convective zone, $\vec{\omega}_{conv}$ is the angular
velocity of the convective zone and $\frac{R_\star}{R_\odot}$ and
$\frac{M_\star}{M_\odot}$ are the radius and mass of the parent star in solar
units respectively.

In addition, angular momentum is exchanged between the radiative core and
convective envelope by mass exchange and by a torque driving the two zones
toward solid body rotation:
\begin{equation}
    \begin{split}
        &\left(\frac{d \vec{J}_{conv}}{d t}\right)_{coup} =
        -\left(\frac{d \vec{J}_{rad}}{d t}\right)_{coup} = \\
        &=
        \frac{1}{\tau_c} 
        \frac{I_{conv} \vec{J}_{rad} - I_{rad} \vec{J}_{conv}} {I_{rad} +
        I_{conv}}
        -
        \frac{2 R_{rad}^2}{3 I_{conv}} \left(\frac{d M_{rad}}{dt}\right)
        \vec{\omega}_{conv}
    \end{split}
\end{equation}
where $\vec{J}_{rad}$ is the angular momentum of the radiative core,
$I_{conv}$ and $I_{rad}$ are the moments of inertia of the convective and
radiative zones respectively, $M_{rad}$ and $R_{rad}$ are the mass and outer
radius of the radiative zone, and $\tau_c$ is a model parameter giving the
timescale on which the core and the envelope converge to solid body rotation.

Finally, we will use YREC tracks \citep{Demarque_et_al_08} for the evolution
of the stellar quantities ($I_{conv}$, $I_{rad}$, $R_\star$, $M_{rad}$ and
$R_{rad}$).

The combined orbital and stellar spin evolution described above was computed
using a more general version of the POET code
\citep{Penev_Zhang_Jackson_14}, which, among other things, allows following
the evolution for systems in which the stellar spin is misaligned with the
orbit.

\subsection{Method}
\label{sec:Q_method}
\begin{figure}[]
    \plotone{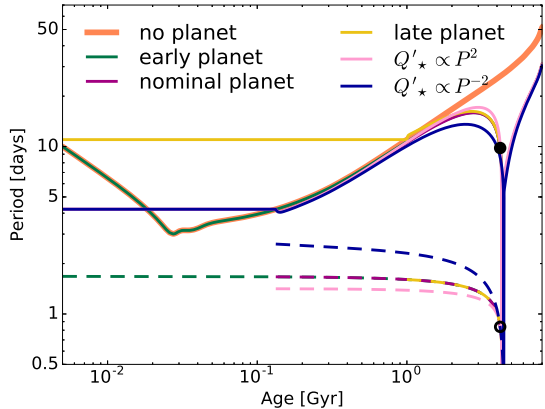}
    \caption[]{
        Example evolution of \hatcur{} spin (solid lines) and \hatcurb{}
        orbital period (dashed lines) using the nominal measured  parameters
        for the system and $\log_{10}(Q'_\star) = 7.3$ at the observed tidal
        frequency of 0.46\,days. The different lines correspond to adding the
        planet at ages 10\,Myr (early planet), 133\,Myrs (nominal planet) and
        1\,Gyr (late planet) as well as two additional assumptions for the
        frequency dependence of $Q'_\star$. The initial orbital period is
        chosen such that the present orbital period is reproduced at the
        present age of 4.2\,Gyrs (open black circle). The initial stellar
        spin at the time the planet is added is that of a star evolving only
        under the influence of angular momentum loss (line labeled no planet)
        due to stellar wind. Regardless of the assumed planet migration age,
        the presently observed stellar spin period is reproduced at the
        present system age (filled black circle), to much better than the
        measurement uncertainty. The different frequency scaling of
        $Q'_\star$ also have a relatively minor effect on the predicted
        stellar spin (both land within 2-sigma of the measured spin period).
    }
    \label{fig:sample_evolutions}
\end{figure}

Given values for all model parameters, in order to fully specify the
evolution of the system, we need to choose appropriate boundary conditions.
Clearly, the observed state of the system provides those, but if we wish to
use the observed stellar spin to constrain $Q'_\star$ we must find
independent spin boundary conditions. Fortunately, rotation periods for stars
in young open clusters have been widely measured. Conveniently, as long as
the stellar spin--down parameters are chosen to reproduce the observed
evolution of stellar spin with age in open clusters, it makes very little
difference which particular cluster we choose to start the evolution from.
This is because for reasonable tidal dissipation rates, only a very tiny
fraction of the orbital evolution occurs in the first few hundred Myrs, and
as a result, the stellar spin evolution hardly differs from that of an
isolated star. This is very fortunate, since our results will not depend on
the formation mechanism of HJs. Whether they form very early through disk
migration, or much later through high--eccentricity migration, will have only
a negligible effect on the final stellar spin. Example evolutions of
\hatcur{}, using the nominal parameters from tables \ref{tab:stellar} and
\ref{tab:planetparam}, adding the planet at ages 10\,Myrs, 133\,Myrs and
1\,Gyr are shown in Fig. \ref{fig:sample_evolutions}. In all cases, the
evolution was started with the spin the star would have if it evolved only
under the influence of angular momentum loss to stellar wind, and the initial
orbital period of the planet was selected to reproduce the currently observed
orbital period at the current age. We can see that, as expected, the effect
of the age at which the planet migrates to its short period orbit on the
stellar spin is utterly negligible compared to the uncertainty of the
measurement. In addition, Fig. \ref{fig:sample_evolutions} also shows that
effect of assuming a frequency dependent tidal dissipation is relatively
small, with even quite steep dependence on period ($Q'_\star \propto P^2$ or
$Q'_\star \propto P^{-2}$) reproducing the currently observed stellar spin to
within 2-sigma of the measured value, as long as $\log_{10}(Q'_\star)=7.3$ at
the observed tidal period for \hatcur{} (0.46 days).

For the constraint derived below, we used the combined spin periods for
M\,50 \citep{Irwin_et_al_09} and the Pleiades \citep{Hartman_et_al_10},
since the two clusters are very close in age, have consistent period
distributions, and together provide a large sample of stars for which the
spin period has been measured. We assumed a starting age of 133\,Myrs for
all evolutions, close to the one estimated for the above clusters.

\ifthenelse{\boolean{emulateapj}}{
    \begin{deluxetable*}{lcccccc}[b!] }{
        \begin{deluxetable}{lcccccc}[b!] 
}
    \tablewidth{0pc}
    \tablecaption{
    The sets of assumptions for which constraints on $\log_{10}(Q'_\star)$
    were derived and the results for each system.
    \label{tab:Q_models}
    }
    \tablehead{
        \colhead{Name} &
        \colhead{$K$} &
        \colhead{$\tau_c$} &
        \colhead{$\omega_{sat}$} &
        \colhead{Initial spin} &
        \colhead{\hatcur{} Constraint} &
        \colhead{WASP-19 Constraint} \\
        &
        \colhead{($\frac{M_\odot R_\odot^2 \mathrm{day}^2}
                        {(\mathrm{rad}^2 \mathrm{Gyr})}$)} &
        \colhead{(Myr)} &
        \colhead{(rad/day)} &
        &
        \colhead{68.2\% Confidence Interval} &
        \colhead{68.2\% Confidence Interval} \\
    }
    \startdata
    \textbf{nominal}\tablenotemark{a} & $ 0.17 $ & $ 10 $ & $ 2.45 $ & prograde & 7.2 --- 7.4 & 6.5 --- 6.9 \\
\textbf{retro}\tablenotemark{a} & $ 0.17 $ & $ 10 $ & $ 2.45 $ & retrograde & 6.8 --- 7.1 & 6.2 --- 6.6 \\
$K = 0.11333$\tablenotemark{b} & $ 0.11333 $  & $ 10 $ & $ 2.45 $ & prograde & 7.3 --- 7.6 & 6.6 --- 7.1 \\
$K = 0.22666$\tablenotemark{b} & $ 0.22666 $  & $ 10 $ & $ 2.45 $ & prograde & 7.1 --- 7.3 & 6.4 --- 6.8 \\
$\tau_c = 1$\tablenotemark{b} & $ 0.17 $ & $ 1 $ & $ 2.45 $ & prograde & 7.0 --- 7.3 & 6.3 --- 6.8 \\
$\tau_c = 25$\tablenotemark{b} & $ 0.17 $ & $ 25 $ & $ 2.45 $ & prograde & 7.3 --- 7.6 & 6.6 --- 7.0 \\
$\omega_{sat} = 1.225$\tablenotemark{b} & $ 0.17 $ & $ 10 $ & $ 1.225 $ & prograde & 7.2 --- 7.4 & 6.5 --- 6.9 \\
$\omega_{sat} = 4.9$\tablenotemark{b} & $ 0.17 $ & $ 10 $ & $ 4.9 $ & prograde & 7.2 --- 7.4 & 6.5 --- 6.8 \\
 [-1.5ex]
    \enddata
    \tablenotetext{a}{
        The values of $K$, $\tau_c$ and $\omega_{sat}$ used for these models
        are best fit values to observations of stellar spin in open clusters
        of various ages.
    }
    \tablenotetext{b}{
        The changes in stellar angular momentum loss parameters used in these
        models do not represent actual uncertainties, but are in fact much
        larger. All of these models are in clear conflict with observations.
        The particular values used were chosen to demonstrate the (lack of)
        sensitivity of the results to each parameter separately.
    }
\ifthenelse{\boolean{emulateapj}}{
    \end{deluxetable*}
}{ 
    \end{deluxetable} 
}

In order to constrain the value of the tidal dissipation parameter $Q'_\star$
defined above, fully accounting for the posterior distributions of the
measured \hatcur{} system properties, we will follow the following procedure: 
\begin{enumerate}
    \item select a random step from the converged DEMCMC chain, thus getting
        values for the present age of the \hatcur{} system as well as the
        stellar and planetary masses and the stellar radius.
    \item Randomly select one of the stars from the Pleiades/M\,50 with
        measured rotation period that has a mass within $0.1M_\odot$ of the
        randomly selected stellar mass above and use its spin period as the
        initial spin for the calculated evolution.
    \item Select a random value for $log_{10}(Q'_\star)$ from a uniform
        distribution in the range $(5, 9)$.
    \item Find an initial orbital period, such that starting the evolution at
        an age of 133\,Myrs with the above parameters and evolving to the
        randomly selected present system age, results in the observed orbital
        period (the comparatively tiny uncertainty in the current orbital
        period is ignored).
    \item Assume a normal distribution for the measured stellar spin period
        at the present age and evaluate the distribution at the resulting
        stellar spin period with the above evolution to get $p(Q'_\star)$.
\end{enumerate}
Repeating the above steps multiple times allows us to build a cumulative
distribution function (CDF) for $log_{10}(Q'_\star)$ by summing up all
$p(Q'_\star)$ values up to a particular $log_{10}(Q'_\star)$. The number of
iterations was chosen such that doubling their number did not result in
significant changes in the CDF.

Finally, the entire procedure was repeated for a number of assumptions about
the parameters of the spin model in order to investigate the sensitivity of
the constraint to these parameters. In addition, even though planets around
stars with surface convective zones appear to be well aligned with their host
star's spin, it is possible that they form with a wide range of obliquities,
which then decay on a timescale short compared to the tidal orbital decay for
typical planets, but it may not be short compared to the orbital decay for
\hatcur{}. In order to investigate the impact this could have on the results,
we also considered the most extreme possible case, of starting the star
spinning in exactly the opposite direction to the orbit and evolving to a
presently assumed prograde state. The particular set of parameters considered
is given in table \ref{tab:Q_models}. The ``nominal'' and ``retrograde''
models use the parameters for the stellar spin evolution which best fit the
observed spin periods of open clusters \citep{Irwin_et_al_07}. An important
point to note is that the change in parameter values away from the nominal
model, for the other cases considered, do not represent actual uncertainties.
In fact, all of these changes are in dramatic conflict with observations,
demonstrating that very large changes in the models are required to make
appreciable changes to the inferred $Q'_\star$ constraint. A more appropriate
treatment, which accounts properly for the shifts in the model parameters
allowed by the cluster data, is beyond the scope of this paper, but the range
of models considered demonstrates the robustness of the results presented
here.

In addition to \hatcur{}, we carried out the steps outlined in the previous
section for WASP-19. This is another one of the three planetary systems whose
measured semimajor axis to stellar radius ratio is consistent with being the
smallest, and hence can be expected to have its host star spun up due to
tidal dissipation. Indeed, it also seems to be spinning faster than expected
for its age. In fact, \citet{TregloanReed_Southworth_Tappert_13} observed the
planet transiting in front of, what appears to be the same star spot, on two
consecutive nights, which allowed them to measure WASP-19's spin period to be
$11.76\pm0.09$\,days, while the discovery paper \citep{Hebb_et_al_10} quoted
a photometrically detected rotation period of $10.5\pm0.2$\,days.  Neither of
these periods is consistent with the isochronal constraint that the system is
older than 1\,Gyr \citep{Hebb_et_al_10}.

In order to make the results from \hatcur{} and WASP-19 as comparable as
possible, we used the same set of isochrones and the same fitting procedure
to derive an isochronal age for WASP-19 of $8\pm3$\,Gyr. Further, both stars
have masses very close to solar, which means we do not need to worry about
dependences of the various model parameters on the stellar mass. Finally, a
proper DEMCMC fit to the WASP-19 observations is not available, so unlike for
\hatcur{}, we simply assume the relevant parameters for WASP-19 from the
literature and use a Normal distribution with the quoted uncertainties. The
particular values we employed were taken from
\citet{TregloanReed_Southworth_Tappert_13}, and are consistent with the rest
of the literature: $M_\star=0.904\pm0.045 M_\odot$, $R_\star=1.004\pm0.018
R_\odot$, $M_{pl}=1.114\pm0.04 M_J$, and we adopted the
\citet{TregloanReed_Southworth_Tappert_13} stellar spin period of
$11.76\pm0.09$\,days and orbital period of $P=0.788840\pm0.0000003$\,days.

\subsection{Results}
\label{sec:Q_results}
\begin{figure}[]
    \plotone{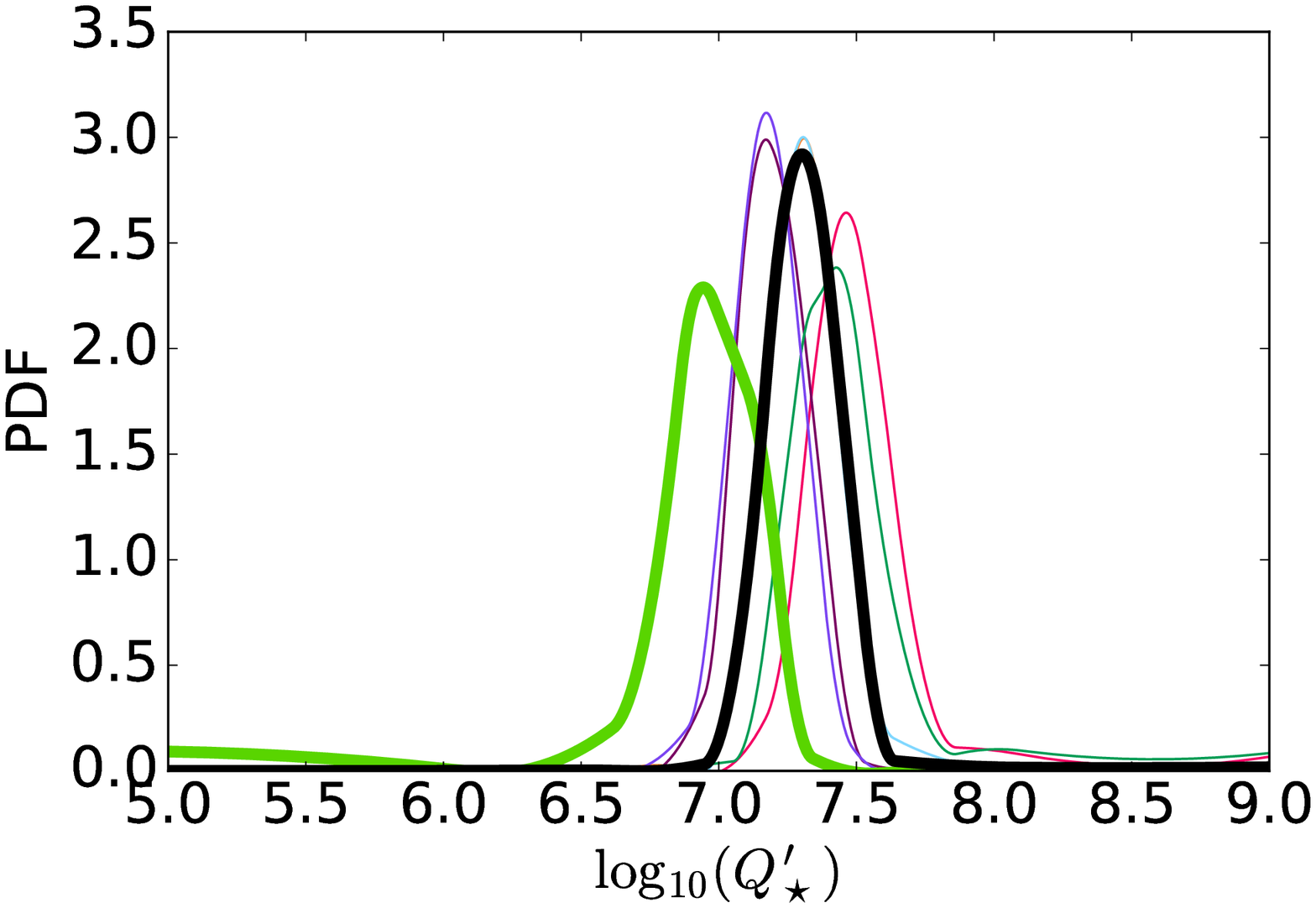}
    \plotone{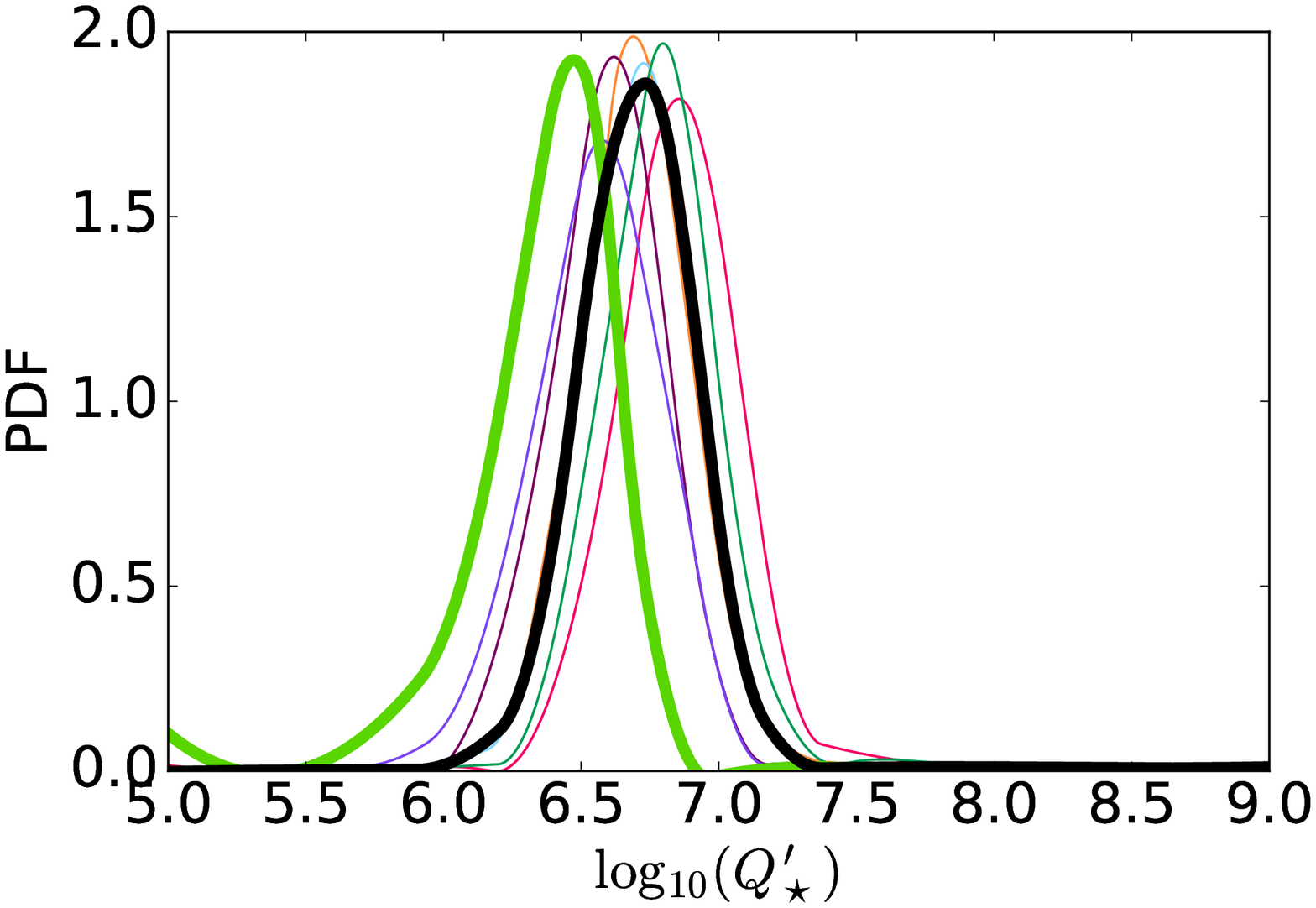}
    \plotone{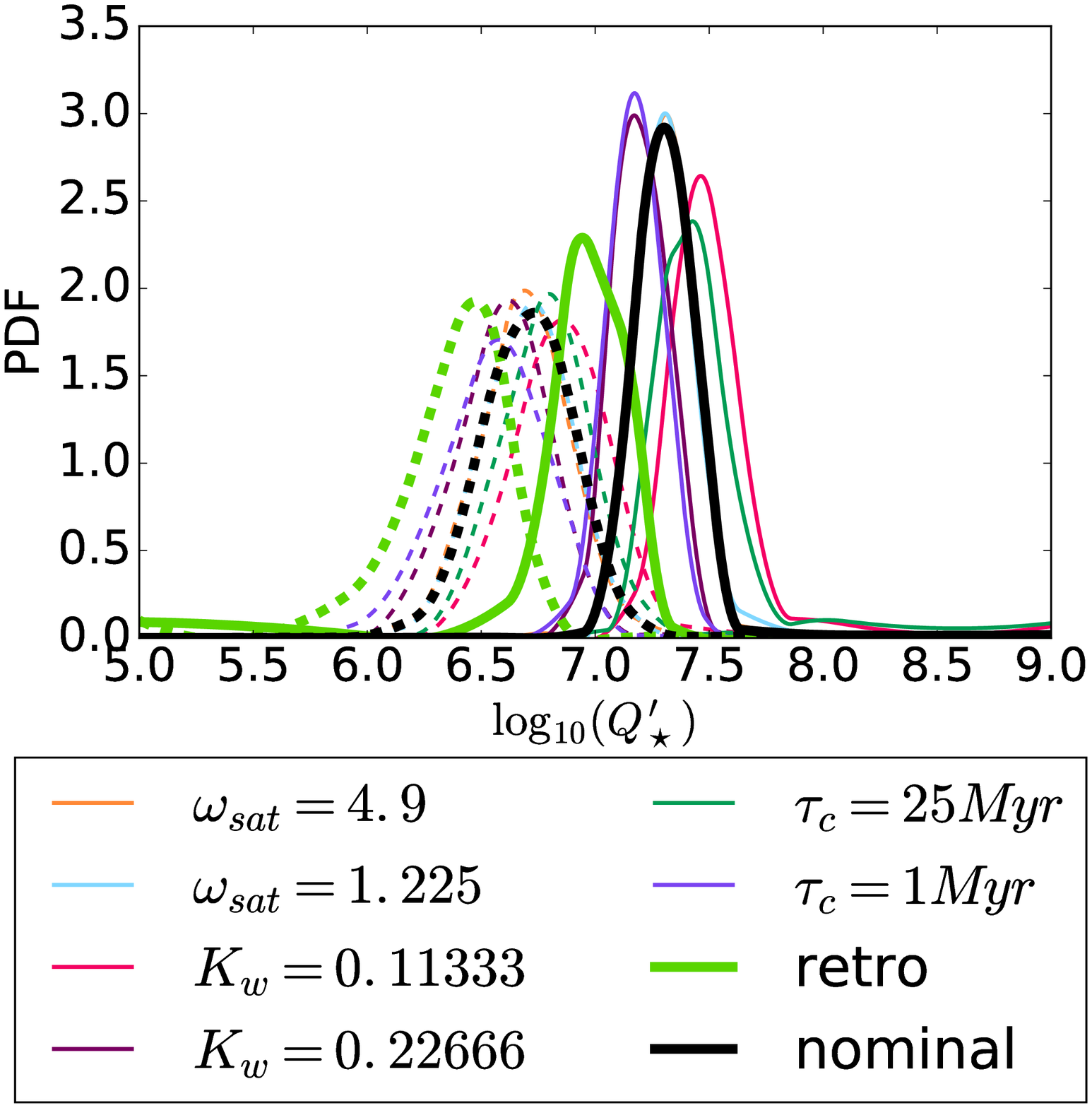}
    \caption[]{
        Top: The PDF of $\log_{10}(Q'_\star)$ from the \hatcur{} system
        parameters. The various lines correspond to the models from
        table~\ref{tab:Q_models} with the models for which the angular momentum
        loss parameters match the best fit of the stellar rotation rates in
        open clusters
        plotted with thicker lines. Middle: The same as the first panel, but
        using WASP-19 system parameters. Bottom: All the curves from the
        previous two plots together with \hatcur{} plotted with solid lines,
        and WAPS-19 with dashed.
    }
    \label{fig:lgQ_pdfs}
\end{figure}

In order to generate plots of the probability density functions (PDF) derived
by the procedure described above, we fit a smoothing bicubic spline to the
cumulative distribution with a tiny amount of smoothing in order to suppress
numerical oscillations when taking the derivative. Fig.\,\ref{fig:lgQ_pdfs}
shows the PDF derived for $\log_{10}(Q'_\star)$ for \hatcur{} and WASP-19 with
the various models of table~\ref{tab:Q_models}. The constraints obtained for
$Q'_\star$ are given in the last column of that table. The confidence interval
was derived by evaluating the inverse cumulative distribution function for
$\log_{10}(Q'_\star)$ at 15.87\% and at 84.13\%.

Since most of the orbital decay happens at late times when the star is
evolving only very slowly on the main sequence, it is a very good
approximation to assume a non--evolving star with the present properties in
the last Gyr or so of the evolution. As a result, as long as the star is
started with the spin predicted by angular momentum loss in the absence of a
planet, the results are only very slightly sensitive to the exact stellar
evolution models used.  In particular this means that the exact stellar age
determined by matching the evolution models to the present star, has only a
very small effect on the results.

Clearly, parametrizing tidal dissipation by a single number ($Q'_\star$ in
our case) is a gross oversimplification of the physics involved. In reality
$Q'_\star$ should depend on the stellar mass, the tidal frequency, and the
stellar spin. This can affect the results in two ways: first, it could be one
way to explain the different results obtained for the two systems, and
second, even for a single system, the spin of the star and the tidal
frequency evolve, thus different tidal dissipation will operate at different
times during the system's past. However, for the two planetary systems
considered all these parameters are currently almost identical. Further, due
to the strong dependence of the rate of orbital decay on the planet--star
separation, and the fact that angular momentum loss is faster for faster
spinning stars, only the most recent part of the evolution of the systems
matters (as demonstrated in Fig. \ref{fig:sample_evolutions}). 

So even though the past spin histories of the two stars may have been
somewhat different (due to the different planetary masses), this has a
relatively small impact on the results. In addition, since the evolution is
dominated by the latest stages, strictly speaking, the constraints derived
here give the tidal quality factor for parameters close to the currently
observed ones (a stellar mass of approximately a $1M_\odot$, for orbital
    periods of approximately 0.8 days and for stellar spin periods of about
10 days). Finally, this also means that the formation mechanism for the
planets is irrelevant for the derived constraints. While it is true that
starting the orbital evolution later, if planet migration is delayed, can
decrease the amount of angular momentum added to the star, this is a totally
negligible effect (see Fig. \ref{fig:sample_evolutions}).

Disentangling the dependence of the tidal dissipation on some of
these quantities may be possible by performing similar analysis on a larger
number of exoplanet systems, ideally all currently known extremely short
period ones. In addition, orbital circularization and spin synchronization in
open cluster binaries is able to probe much longer orbital periods than is
feasible with extrasolar planets.

\section{Discussion}
\label{sec:discussion}
\hatcur{} is an extreme short--period planet which is among the best targets
for testing theories of planet--star interactions. In fact, the host star,
like a number of other extremely short period giant--planet hosts (e.g.
    WASP-19 above, WASP-103 \citep{Gillon_et_al_14}, OGLE-TR-113
\citep{Bouchy_et_al_04}) appears to be spinning too fast for its age.
\hatcur{} is the best system to--date for constraining the stellar tidal
dissipation by assuming that the extra stellar angular momentum was delivered
by tidal decay of the orbit. In fact, we applied this method to the two
exoplanet systems whose host stars should have been spun up the most, and
which have very similar properties, to derive tight constraints on the
stellar tidal quality factor at least in the regime applicable to those
systems. In fact, if both of these planets are assumed to have formed in
orbits well aligned with their parent star's spin, there is only a very
narrow range around $\log_{10}(Q'_\star)=7$ for which the present spin period
of both stars is at least marginally consistent with the expected degree of
spin--up. This tight constraint will also apply if planets form with a wide
range of initial obliquities, but are quickly re-aligned by some process
which operates on timescales short compared to the orbital decay. On the
other hand, if planets are assumed to form with a wide range of obliquities,
and if at least for the extremely short periods of \hatcur{} and WASP-19, the
timescale for orbital decay is shorter than any processes which tend to align
the orbit with the stellar equator, it is plausible that WASP-19 started out
in a well aligned orbit, while \hatcur{} was significantly misaligned in
which case, $6.5 < \log_{10}(Q'_\star) < 7$.  Clearly, a more systematic
effort to analyze all suitable exoplanet systems and properly account for the
stellar angular momentum loss uncertainties is bound to yield very meaningful
constraints on the stellar tidal dissipation, as well how it changes with
various system properties.

These constraints do not match the recently suggested detection of orbital
decay in WASP-12 \citep{Maciejewski_et_al_16}, which would correspond to a
tidal quality factor of $Q'_\star = 2.5\times10^5$. However, the authors of
that study point out that at present the observed period change is still
marginally consistent with apsidal precession. Further, as we pointed out
above, the tidal quality factor is not expected to be the same across
different systems, and WASP-12 differs from both \hatcur{} and WASP-19 in
several important respects: it has a hotter star, with only a minimal surface
convective zone, and it appears to be spinning significantly slower.  Both of
these properties are expected to impact the tidal dissipation. The same
measurement is also within reach for \hatcurb{}. For example, after 28
years, the time of arrival of \hatcurb{} transits will have shifted by 60\,s
if $Q'_\star=10^7$ due to tidal orbital decay, thus making it feasible to
measure.

As we argued in \refsec{comparison}, extremely short period planets like
\hatcur{} provide a fantastic laboratory to test a range of interactions
between the planet and the star, and hence, expanding this sample is
extremely valuable for the study of extrasolar planets.


\acknowledgements 

\paragraph{Acknowledgements}
Development of the HATSouth project was funded by NSF MRI grant
NSF/AST-0723074, operations have been supported by NASA grants
NNX09AB29G and NNX12AH91H, and follow-up observations received partial
support from grant NSF/AST-1108686.
K.P.\ acknowledges support from NASA grants NNX13AQ62G and NNG14FC03C.
G.B.\ acknowledges support from the David and Lucile Packard Foundation, from NASA grants NNX13AJ15G, NNX14AF87G and NNX13AQ62G.
J.H.\ acknowledges support from NASA grants NNX13AJ15G and NNX14AF87G.
R.B.\ and N.E.\ are supported by CONICYT-PCHA/Doctorado Nacional.
A.J.\ acknowledges support from FONDECYT project 1130857, BASAL CATA
PFB-06, and from the Ministry of Economy, Development, and Tourism's Millenium Science Initiative through grant IC120009, awarded to the Millenium Institute of Astrophysics, MAS.\ R.B.\ and N.E.\ acknowledge additional support from the Ministry of Economy, Development, and Tourism's Millenium Science Initiative through grant IC120009, awarded to the Millenium Institute of Astrophysics, MAS.\ V.S.\ acknowledges support from BASAL CATA PFB-06. 
This paper uses observations obtained with facilities of the Las
Cumbres Observatory Global Telescope.
Work at the Australian National University is supported by ARC Laureate
Fellowship Grant FL0992131.
We acknowledge the use of the AAVSO Photometric All-Sky Survey (APASS),
funded by the Robert Martin Ayers Sciences Fund, and the SIMBAD
database, operated at CDS, Strasbourg, France.
Operations at the MPG~2.2\,m Telescope are jointly performed by the
Max Planck Gesellschaft and the European Southern Observatory.
%
G.~B.~wishes to thank the warm hospitality of Ad\'ele and Joachim Cranz
at the farm Isabis, supporting the operations and service missions of
HATSouth.

\clearpage
\bibliographystyle{apj}
\bibliography{hatsbib,discussionbib}

\begin{thebibliography}{43}
\expandafter\ifx\csname natexlab\endcsname\relax\def\natexlab#1{#1}\fi

\bibitem[{{Ag{\"u}eros} {et~al.}(2011){Ag{\"u}eros}, {Covey}, {Lemonias},
  {Law}, {Kraus}, {Batalha}, {Bloom}, {Cenko}, {Kasliwal}, {Kulkarni},
  {Nugent}, {Ofek}, {Poznanski}, \& {Quimby}}]{Agueros_et_al_11}
{Ag{\"u}eros}, M.~A., {Covey}, K.~R., {Lemonias}, J.~J., {et~al.} 2011, \apj,
  740, 110

\bibitem[{{Albrecht} {et~al.}(2012){Albrecht}, {Winn}, {Johnson}, {Howard},
  {Marcy}, {Butler}, {Arriagada}, {Crane}, {Shectman}, {Thompson}, {Hirano},
  {Bakos}, \& {Hartman}}]{Albrecht_et_al_12}
{Albrecht}, S., {Winn}, J.~N., {Johnson}, J.~A., {et~al.} 2012, \apj, 757, 18

\bibitem[{{Amard} {et~al.}(2016){Amard}, {Palacios}, {Charbonnel}, {Gallet}, \&
  {Bouvier}}]{Amard_et_al_16}
{Amard}, L., {Palacios}, A., {Charbonnel}, C., {Gallet}, F., \& {Bouvier}, J.
  2016, \aap, 587, A105

\bibitem[{{Bakos} {et~al.}(2010){Bakos}, {Torres}, {P{\'a}l}, {Hartman},
  {Kov{\'a}cs}, {Noyes}, {Latham}, {Sasselov}, {Sip{\H o}cz}, {Esquerdo},
  {Fischer}, {Johnson}, {Marcy}, {Butler}, {Isaacson}, {Howard}, {Vogt},
  {Kov{\'a}cs}, {Fernandez}, {Mo{\'o}r}, {Stefanik}, {L{\'a}z{\'a}r}, {Papp},
  \& {S{\'a}ri}}]{bakos:2010:hat11}
{Bakos}, G.~{\'A}., {Torres}, G., {P{\'a}l}, A., {et~al.} 2010, \apj, 710, 1724

\bibitem[{{Bakos} {et~al.}(2013){Bakos}, {Csubry}, {Penev}, {Bayliss},
  {Jord{\'a}n}, {Afonso}, {Hartman}, {Henning}, {Kov{\'a}cs}, {Noyes},
  {B{\'e}ky}, {Suc}, {Cs{\'a}k}, {Rabus}, {L{\'a}z{\'a}r}, {Papp}, {S{\'a}ri},
  {Conroy}, {Zhou}, {Sackett}, {Schmidt}, {Mancini}, {Sasselov}, \&
  {Ueltzhoeffer}}]{bakos:2013:hatsouth}
{Bakos}, G.~{\'A}., {Csubry}, Z., {Penev}, K., {et~al.} 2013, \pasp, 125, 154

\bibitem[{{Barnes} {et~al.}(2016){Barnes}, {Weingrill}, {Fritzewski}, \&
  {Strassmeier}}]{Barnes_et_al_16}
{Barnes}, S.~A., {Weingrill}, J., {Fritzewski}, D., \& {Strassmeier}, K.~G.
  2016, ArXiv e-prints, 1603.09179

\bibitem[{{Bayliss} {et~al.}(2013){Bayliss}, {Zhou}, {Penev}, {Bakos},
  {Hartman}, {Jord{\'a}n}, {Mancini}, {Mohler-Fischer}, {Suc}, {Rabus},
  {B{\'e}ky}, {Csubry}, {Buchhave}, {Henning}, {Nikolov}, {Cs{\'a}k}, {Brahm},
  {Espinoza}, {Noyes}, {Schmidt}, {Conroy}, {Wright}, {Tinney}, {Addison},
  {Sackett}, {Sasselov}, {L{\'a}z{\'a}r}, {Papp}, \&
  {S{\'a}ri}}]{bayliss:2013:hats3}
{Bayliss}, D., {Zhou}, G., {Penev}, K., {et~al.} 2013, \aj, 146, 113

\bibitem[{{Bouchy} {et~al.}(2004){Bouchy}, {Pont}, {Santos}, {Melo}, {Mayor},
  {Queloz}, \& {Udry}}]{Bouchy_et_al_04}
{Bouchy}, F., {Pont}, F., {Santos}, N.~C., {et~al.} 2004, \aap, 421, L13

\bibitem[{{Brahm et. al.}(2016)}]{brahm:2016:zaspe}
{Brahm et. al.}, . 2016

\bibitem[{{Cardelli} {et~al.}(1989){Cardelli}, {Clayton}, \&
  {Mathis}}]{cardelli:1989}
{Cardelli}, J.~A., {Clayton}, G.~C., \& {Mathis}, J.~S. 1989, \apj, 345, 245

\bibitem[{{Claret}(2004)}]{claret:2004}
{Claret}, A. 2004, \aap, 428, 1001

\bibitem[{{Dawson} \& {Murray-Clay}(2013)}]{Dawson_MurrayClay_13}
{Dawson}, R.~I., \& {Murray-Clay}, R.~A. 2013, \apjl, 767, L24

\bibitem[{{Delorme} {et~al.}(2011){Delorme}, {Collier Cameron}, {Hebb},
  {Rostron}, {Lister}, {Norton}, {Pollacco}, \& {West}}]{Delorme_et_al_11}
{Delorme}, P., {Collier Cameron}, A., {Hebb}, L., {et~al.} 2011, \mnras, 413,
  2218

\bibitem[{{Demarque} {et~al.}(2008){Demarque}, {Guenther}, {Li}, {Mazumdar}, \&
  {Straka}}]{Demarque_et_al_08}
{Demarque}, P., {Guenther}, D.~B., {Li}, L.~H., {Mazumdar}, A., \& {Straka},
  C.~W. 2008, \apss, 316, 31

\bibitem[{{Dopita} {et~al.}(2007){Dopita}, {Hart}, {McGregor}, {Oates},
  {Bloxham}, \& {Jones}}]{dopita:2007}
{Dopita}, M., {Hart}, J., {McGregor}, P., {et~al.} 2007, \apss, 310, 255

\bibitem[{{Fossati} {et~al.}(2010){Fossati}, {Haswell}, {Froning}, {Hebb},
  {Holmes}, {Kolb}, {Helling}, {Carter}, {Wheatley}, {Collier Cameron},
  {Loeillet}, {Pollacco}, {Street}, {Stempels}, {Simpson}, {Udry}, {Joshi},
  {West}, {Skillen}, \& {Wilson}}]{Fossati_et_al_10}
{Fossati}, L., {Haswell}, C.~A., {Froning}, C.~S., {et~al.} 2010, \apjl, 714,
  L222

\bibitem[{{Fressin} {et~al.}(2013){Fressin}, {Torres}, {Charbonneau}, {Bryson},
  {Christiansen}, {Dressing}, {Jenkins}, {Walkowicz}, \&
  {Batalha}}]{Fressin_et_al_13}
{Fressin}, F., {Torres}, G., {Charbonneau}, D., {et~al.} 2013, \apj, 766, 81

\bibitem[{{Gallet} \& {Bouvier}(2015)}]{Gallet_Bouvier_15}
{Gallet}, F., \& {Bouvier}, J. 2015, \aap, 577, A98

\bibitem[{{Gillon} {et~al.}(2014){Gillon}, {Anderson}, {Collier-Cameron},
  {Delrez}, {Hellier}, {Jehin}, {Lendl}, {Maxted}, {Pepe}, {Pollacco},
  {Queloz}, {S{\'e}gransan}, {Smith}, {Smalley}, {Southworth}, {Triaud},
  {Udry}, {Van Grootel}, \& {West}}]{Gillon_et_al_14}
{Gillon}, M., {Anderson}, D.~R., {Collier-Cameron}, A., {et~al.} 2014, \aap,
  562, L3

\bibitem[{{Ginzburg} \& {Sari}(2015)}]{Ginzburg_Sari_15}
{Ginzburg}, S., \& {Sari}, R. 2015, \apj, 803, 111

\bibitem[{{Hansen} \& {Barman}(2007)}]{hansen:2007}
{Hansen}, B.~M.~S., \& {Barman}, T. 2007, \apj, 671, 861

\bibitem[{{Hartman} {et~al.}(2010){Hartman}, {Bakos}, {Kov{\'a}cs}, \&
  {Noyes}}]{Hartman_et_al_10}
{Hartman}, J.~D., {Bakos}, G.~{\'A}., {Kov{\'a}cs}, G., \& {Noyes}, R.~W. 2010,
  \mnras, 408, 475

\bibitem[{{Hartman} {et~al.}(2009){Hartman}, {Gaudi}, {Pinsonneault}, {Stanek},
  {Holman}, {McLeod}, {Meibom}, {Barranco}, \& {Kalirai}}]{Hartman_et_al_09}
{Hartman}, J.~D., {Gaudi}, B.~S., {Pinsonneault}, M.~H., {et~al.} 2009, \apj,
  691, 342

\bibitem[{{Hartman} {et~al.}(2012){Hartman}, {Bakos}, {B{\'e}ky}, {Torres},
  {Latham}, {Csubry}, {Penev}, {Shporer}, {Fulton}, {Buchhave}, {Johnson},
  {Howard}, {Marcy}, {Fischer}, {Kov{\'a}cs}, {Noyes}, {Esquerdo}, {Everett},
  {Szklen{\'a}r}, {Quinn}, {Bieryla}, {Knox}, {Hinz}, {Sasselov}, {F{\H
  u}r{\'e}sz}, {Stefanik}, {L{\'a}z{\'a}r}, {Papp}, \&
  {S{\'a}ri}}]{hartman:2012:hat39hat41}
{Hartman}, J.~D., {Bakos}, G.~{\'A}., {B{\'e}ky}, B., {et~al.} 2012, \aj, 144,
  139

\bibitem[{{Haswell} {et~al.}(2012){Haswell}, {Fossati}, {Ayres}, {France},
  {Froning}, {Holmes}, {Kolb}, {Busuttil}, {Street}, {Hebb}, {Collier Cameron},
  {Enoch}, {Burwitz}, {Rodriguez}, {West}, {Pollacco}, {Wheatley}, \&
  {Carter}}]{Haswell_et_al_12}
{Haswell}, C.~A., {Fossati}, L., {Ayres}, T., {et~al.} 2012, \apj, 760, 79

\bibitem[{{Hebb} {et~al.}(2010){Hebb}, {Collier-Cameron}, {Triaud}, {Lister},
  {Smalley}, {Maxted}, {Hellier}, {Anderson}, {Pollacco}, {Gillon}, {Queloz},
  {West}, {Bentley}, {Enoch}, {Haswell}, {Horne}, {Mayor}, {Pepe}, {Segransan},
  {Skillen}, {Udry}, \& {Wheatley}}]{Hebb_et_al_10}
{Hebb}, L., {Collier-Cameron}, A., {Triaud}, A.~H.~M.~J., {et~al.} 2010, \apj,
  708, 224

\bibitem[{{Ida} \& {Lin}(2008)}]{Ida_Lin_08}
{Ida}, S., \& {Lin}, D.~N.~C. 2008, \apj, 673, 487

\bibitem[{{Irwin} {et~al.}(2009){Irwin}, {Aigrain}, {Bouvier}, {Hebb},
  {Hodgkin}, {Irwin}, \& {Moraux}}]{Irwin_et_al_09}
{Irwin}, J., {Aigrain}, S., {Bouvier}, J., {et~al.} 2009, \mnras, 392, 1456

\bibitem[{{Irwin} {et~al.}(2007){Irwin}, {Hodgkin}, {Aigrain}, {Hebb},
  {Bouvier}, {Clarke}, {Moraux}, \& {Bramich}}]{Irwin_et_al_07}
{Irwin}, J., {Hodgkin}, S., {Aigrain}, S., {et~al.} 2007, \mnras, 377, 741

\bibitem[{{Jord{\'a}n} {et~al.}(2014){Jord{\'a}n}, {Brahm}, {Bakos}, {Bayliss},
  {Penev}, {Hartman}, {Zhou}, {Mancini}, {Mohler-Fischer}, {Ciceri}, {Sato},
  {Csubry}, {Rabus}, {Suc}, {Espinoza}, {Bhatti}, {Borro}, {Buchhave},
  {Cs{\'a}k}, {Henning}, {Schmidt}, {Tan}, {Noyes}, {B{\'e}ky}, {Butler},
  {Shectman}, {Crane}, {Thompson}, {Williams}, {Martin}, {Contreras},
  {L{\'a}z{\'a}r}, {Papp}, \& {S{\'a}ri}}]{jordan:2014:hats4}
{Jord{\'a}n}, A., {Brahm}, R., {Bakos}, G.~{\'A}., {et~al.} 2014, \aj, 148, 29

\bibitem[{{Kaufer} \& {Pasquini}(1998)}]{kaufer:1998}
{Kaufer}, A., \& {Pasquini}, L. 1998, in Society of Photo-Optical
  Instrumentation Engineers (SPIE) Conference Series, Vol. 3355, Optical
  Astronomical Instrumentation, ed. S.~{D'Odorico}, 844--854

\bibitem[{{Kov{\'a}cs} {et~al.}(2005){Kov{\'a}cs}, {Bakos}, \&
  {Noyes}}]{kovacs:2005:TFA}
{Kov{\'a}cs}, G., {Bakos}, G., \& {Noyes}, R.~W. 2005, \mnras, 356, 557

\bibitem[{{Kov{\'a}cs} {et~al.}(2002){Kov{\'a}cs}, {Zucker}, \&
  {Mazeh}}]{kovacs:2002:BLS}
{Kov{\'a}cs}, G., {Zucker}, S., \& {Mazeh}, T. 2002, \aap, 391, 369

\bibitem[{{Kov{\'a}cs} {et~al.}(2014){Kov{\'a}cs}, {Hartman}, {Bakos}, {Quinn},
  {Penev}, {Latham}, {Bhatti}, {Csubry}, \& {de Val-Borro}}]{Kovacs_et_al_14}
{Kov{\'a}cs}, G., {Hartman}, J.~D., {Bakos}, G.~{\'A}., {et~al.} 2014, \mnras,
  442, 2081

\bibitem[{{Lai}(2012)}]{Lai_12}
{Lai}, D. 2012, \mnras, 423, 486

\bibitem[{{Maciejewski} {et~al.}(2016){Maciejewski}, {Dimitrov},
  {Fern{\'a}ndez}, {Sota}, {Nowak}, {Ohlert}, {Nikolov}, {Bukowiecki}, {Hinse},
  {Pall{\'e}}, {Tingley}, {Kjurkchieva}, {Lee}, \&
  {Lee}}]{Maciejewski_et_al_16}
{Maciejewski}, G., {Dimitrov}, D., {Fern{\'a}ndez}, M., {et~al.} 2016, \aap,
  588, L6

\bibitem[{{Mandel} \& {Agol}(2002)}]{mandel:2002}
{Mandel}, K., \& {Agol}, E. 2002, \apjl, 580, L171

\bibitem[{{Penev} {et~al.}(2012){Penev}, {Jackson}, {Spada}, \&
  {Thom}}]{Penev_et_al_12}
{Penev}, K., {Jackson}, B., {Spada}, F., \& {Thom}, N. 2012, \apj, 751, 96

\bibitem[{{Penev} {et~al.}(2014){Penev}, {Zhang}, \&
  {Jackson}}]{Penev_Zhang_Jackson_14}
{Penev}, K., {Zhang}, M., \& {Jackson}, B. 2014, \pasp, 126, 553

\bibitem[{{Penev} {et~al.}(2013){Penev}, {Bakos}, {Bayliss}, {Jord{\'a}n},
  {Mohler}, {Zhou}, {Suc}, {Rabus}, {Hartman}, {Mancini}, {B{\'e}ky}, {Csubry},
  {Buchhave}, {Henning}, {Nikolov}, {Cs{\'a}k}, {Brahm}, {Espinoza}, {Conroy},
  {Noyes}, {Sasselov}, {Schmidt}, {Wright}, {Tinney}, {Addison},
  {L{\'a}z{\'a}r}, {Papp}, \& {S{\'a}ri}}]{penev:2013:hats1}
{Penev}, K., {Bakos}, G.~{\'A}., {Bayliss}, D., {et~al.} 2013, \aj, 145, 5

\bibitem[{{ter Braak}(2006)}]{terbraak:2006}
{ter Braak}, C.~J.~F. 2006, Statistics and Computing, 16, 239

\bibitem[{{Tregloan-Reed} {et~al.}(2013){Tregloan-Reed}, {Southworth}, \&
  {Tappert}}]{TregloanReed_Southworth_Tappert_13}
{Tregloan-Reed}, J., {Southworth}, J., \& {Tappert}, C. 2013, \mnras, 428, 3671

\bibitem[{{Yi} {et~al.}(2001){Yi}, {Demarque}, {Kim}, {Lee}, {Ree}, {Lejeune},
  \& {Barnes}}]{yi:2001}
{Yi}, S., {Demarque}, P., {Kim}, Y.-C., {et~al.} 2001, \apjs, 136, 417

\end{thebibliography}

\end{document}